\documentclass[a4paper,12pt]{article}
\usepackage{jheppub,esint,shuffle,psfrag}
 \usepackage[utf8]{inputenc}

\usepackage{esint} 
\usepackage{breqn}
\usepackage[utf8]{inputenc}

\def \tr {\mathop{\rm tr}\nolimits}
\def \Im {\mathop{\rm Im}\nolimits}
\def \Re {\mathop{\rm Re}\nolimits}

\def \e  {\mathop{\rm e}\nolimits}

\newcommand\lr[1]{{\left({#1}\right)}}

\newcommand \vev [1] {\langle{#1}\rangle}

\newcommand\re[1]{(\ref{#1})}

\def \qqqquad {\qquad\qquad}

\def\Xint#1{\mathchoice
   {\XXint\displaystyle\textstyle{#1}}%
   {\XXint\textstyle\scriptstyle{#1}}%
   {\XXint\scriptstyle\scriptscriptstyle{#1}}%
   {\XXint\scriptscriptstyle\scriptscriptstyle{#1}}%
   \!\int}
\def\XXint#1#2#3{{\setbox0=\hbox{$#1{#2#3}{\int}$}
     \vcenter{\hbox{$#2#3$}}\kern-.5\wd0}}

\def\dashint{\Xint-}

\def\numberbysection{\@addtoreset{equation}{section}
                     \def\theequation{\thesection.\arabic{equation}}}

\title{Exact scattering amplitudes in conformal fishnet theory}
 
\author {G.P.~Korchemsky}
 \affiliation {Institut de Physique Th\'eorique\footnote{Unit\'e Mixte de Recherche 3681 du CNRS}, Universit\'e Paris Saclay, CNRS, CEA, 91191 Gif-sur-Yvette}
\preprint{  \parbox[t]{28mm}{IPhT--T18/150}}

 \abstract
{
We compute the leading-color contribution to four-particle scattering amplitude in four-dimensional conformal fishnet theory that arises as a special limit
of $\gamma$-deformed $\mathcal N=4$ SYM. We show that the single-trace partial amplitude is protected from quantum corrections whereas the double-trace partial amplitude is a nontrivial infrared finite function of the ratio of Mandelstam invariants. Applying the Lehmann--Symanzik--Zimmerman reduction procedure to the 
known expression of a four-point correlation function in the fishnet theory, we derive a new representation for this function that is valid for arbitrary coupling. We use this representation to find the asymptotic behavior of the double-trace amplitude in the high-energy limit and to compute the corresponding exact Regge trajectories.
We verify that at weak coupling the expressions obtained are in agreement with an explicit five-loop calculation. 
 }

\begin{document}

\maketitle

\flushbottom

\section{Introduction}

Remarkable progress has recently been achieved in understanding the properties of scattering amplitudes in four-dimensional gauge theories and, 
most notably, in maximally supersymmetric Yang-Mills theory ($\mathcal N=4$ SYM) (see Ref.~\cite{Amplitudes2018} for recent progress). The latter theory is believed to be integrable in the planar limit  \cite{Beisert:2010jr} and scattering 
amplitudes can be used as a powerful tool for uncovering its hidden symmetries. 

Exploiting the symmetries of scattering amplitudes we encounter a few obstacles. First of all, scattering amplitudes suffer from infrared divergences and require introducing a regulator (e.g. dimensional regularization). This  
breaks some of the symmetries, like conformal symmetry and its supersymmetric extension. Finding the corresponding anomalies proves to be a nontrivial task \cite{Drummond:2007au,CaronHuot:2011kk,Bullimore:2011kg,Chicherin:2017bxc}. Secondly, with the exception of four and five particles,
generic scattering amplitudes in planar $\mathcal N=4$ SYM do not admit a closed analytical representation for an arbitrary 't Hooft coupling. 
The former amplitudes are fixed unambiguously by anomalous Ward identities corresponding to the dual conformal symmetry \cite{Drummond:2007au}. 
 Their finite part depends on the coupling constant through the cusp anomalous dimension and is given by the BDS ansatz  \cite{Bern:2005iz}.
   
Beyond the planar limit, 
the scattering amplitudes in $\mathcal N=4$ SYM can be expanded over an appropriately chosen basis of multi-trace color tensors built from generators of the $SU(N)$ gauge group in the fundamental representation. In the simplest case of four-particle amplitude we have
\begin{align}\label{color-ordered}
\mathcal A  = (2\pi)^4 \delta^{(4)}\big(\sum_i p_i\big) \left[N \tr(T^{a_1} T^{a_2} T^{a_3} T^{a_4})A^{\rm (s)} + \tr(T^{a_1} T^{a_2})  \tr(T^{a_3} T^{a_4})  A^{\rm (d)}\right]
+ \text{perm}\,,
\end{align}
where the scattered particles carry on-shell momenta  $p_i$ (taken to be incoming) and  the color charge $T^{a_i}$ (normalized as $\tr(T^a T^b)=\delta^{ab}$). In the case of identical particles, by virtue of Bose symmetry 
the expression on the right-hand side of \re{color-ordered} contains additional terms denoted by `perm' with momenta and color indices exchanged.  The color-ordered partial amplitudes $A^{\rm (s)}(p_i)$ and $A^{\rm (d)}(p_i)$ describe single- and double-trace contributions, respectively.
 In the planar limit, the leading contribution  comes from $A^{\rm (s)}(p_i)$ and it exhibits remarkable properties \cite{Drummond:2008vq}. It remains unclear whether some of these properties survive beyond the planar limit, see  
Refs.~\cite{Bern:2017gdk,Bern:2018oao,Ben-Israel:2018ckc,Chicherin:2018wes} for a recent development.

In this paper, we study scattering amplitudes in a nontrivial four-dimensional conformal theory, the so called fishnet theory.
This theory is closely related to $\mathcal N=4$ SYM and, most importantly, it allows us to avoid some of the difficulties mentioned above.
It is described by the Lagrangian proposed in Ref.~\cite{Gurdogan:2015csr}, 
\begin{align}\label{L0} 
\mathcal L_{\rm cl} = N \tr\Big[ \partial^\mu \bar X \partial_\mu X+\partial^\mu \bar Z \partial_\mu Z 
+ (4\pi \xi)^2 \bar X \bar Z X Z\Big]\,,
\end{align}
where  $X$, $Z$ are complex $N\times N$ traceless matrix scalar fields and $\bar X$, $\bar Z$ denote the conjugated fields. 
 In the planar limit, $N\to\infty$ with $\xi^2$ fixed, correlation functions and scattering amplitudes receive contributions only
from the special class of fishnet Feynman graphs \cite{Zamolodchikov:1980mb} (hence the name of the theory).  Due to the particular CPT noninvariant form of the quartic interaction term in \re{L0}, the theory is nonunitary. As we show below, this leads to a number of unusual properties of the scattering amplitudes. 

The fishnet theory \re{L0} naturally appears in the study of the integrable deformations of maximally supersymmetric Yang-Mills theory. The general $\gamma$-deformed $\mathcal N=4$ SYM theory depends on three deformation parameters \cite{Lunin:2005jy,Frolov:2005dj,Leigh1995}.
The Lagrangian \re{L0} arises in the double scaling limit  in which the Yang-Mills coupling vanishes, $g_{\rm YM}^2\to 0$, and one of the deformation
parameters goes to infinity, $\gamma_3\to i\infty$, in such a way that the product $\xi^2= g_{\rm YM}^2 N \e^{-i\gamma_3}$ remains finite~\cite{Gurdogan:2015csr}. Then, 
all the fields of $\mathcal N=4$ SYM except the two scalars $X$ and $Z$  decouple leading to \re{L0}. Although most of the symmetries (supersymmetry, gauge symmetry)  are broken in this limit~\footnote{The Lagrangian \re{L0} is invariant under global $SU(N)$ and $U(1)\times U(1)$ transformations of the scalar fields, which are remnants of the gauge symmetry and $R-$symmetry of $\mathcal N=4$ SYM, respectively.}, the scattering amplitudes in the fishnet theory are believed to inherit the integrability of $\mathcal N=4$ SYM, at least  in the planar limit \cite{Chicherin:2017cns,Chicherin:2017frs}. 

The Lagrangian \re{L0} is not complete at the quantum level and it should be supplemented by a complete set of
counter-terms \cite{Dymarsky:2005uh,Pomoni:2008de,Fokken:2014soa}. In the planar limit, for  $N\to\infty$ and $\xi^2={\rm fixed}$, they take the form of double-trace dimension-four
operators
\footnote{The contribution of the remaining double-trace counter-terms, like $\tr(X \bar X) \tr(X \bar X)$, is suppressed by a factor of $1/N^2$.}
\begin{align}\notag\label{L-dt}
\mathcal L =\mathcal L_{\rm cl}   {}& +(4\pi)^2 \alpha_1^2 \left[\tr(X^2) \tr(\bar X^2) + \tr(Z^2) \tr(\bar Z^2) \right]
\\[2mm]
{}& -(4\pi)^2\alpha_2^2\left[\tr(XZ) \tr(\bar X\bar Z)+\tr(X\bar Z) \tr(\bar XZ)\right],
\end{align}
where $\alpha_1^2$ and $\alpha_2^2$ are new, induced, coupling constants and the factor of $(4\pi)^2$ is introduced for
convenience. 

The double-trace coupling constants develop nontrivial beta functions and, therefore, the conformal symmetry of the fishnet theory \re{L-dt} is broken \cite{Fokken:2013aea,Sieg:2016vap,Grabner:2017pgm}. Examining the zeros of the beta-functions we find that, in the planar limit, for arbitrary single-trace coupling $\xi^2$, 
the theory has two fixed points  
\begin{align}\label{fixed}
(\alpha_1^2=\alpha_+^2\,,\ \alpha_2^2=\xi^2) \qquad \text{and} \qquad (\alpha_1^2=\alpha_-^2\,,\ \alpha_2^2=\xi^2)\,,
\end{align}
where $\alpha_\pm^2$ is given at weak coupling by  \cite{Grabner:2017pgm}
\begin{align}\label{apm}
\alpha_{\pm}^2=\pm {i\xi^2\over 2} -{\xi^4\over 2} \mp {3i\xi^6\over 4} + \xi^8 \pm {65i \xi^{10}\over 48} -{19\xi^{12}\over 10} + 
O(\xi^{14})\,.
\end{align}
For coupling constants satisfying \re{fixed}, the fishnet theory possesses a conformal symmetry in the planar limit. Notice that the double-trace couplings  $\alpha_+^2$ and $\alpha_-^2$ take complex values for real $\xi^2$ and are related to each other through
$\xi^2\to -\xi^2$. This allows us to restrict the following consideration to one of the fixed points.

The fishnet theory is believed to be integrable at the fixed points \cite{Gurdogan:2015csr,Caetano:2016ydc,Gromov:2017cja,Grabner:2017pgm}. In particular, the
various four-point correlation functions of the shortest scalar operators can be computed exactly.
In this paper, we extend the analysis of Ref.~\cite{Gromov:2018hut} and derive exact expressions for the simplest four-particle scattering amplitudes 
\re{color-ordered} in the conformal fishnet theory \re{L-dt}. We show that the leading large-$N$ contribution to the single- and
double-trace partial amplitudes in \re{color-ordered} are free from infrared and ultraviolet divergences. As a result, $A^{\rm (s)}$ and $A^{\rm (d)}$ are well-defined in four dimensions and respect the exact conformal symmetry. 

We demonstrate that 
the single-trace contribution $A^{\rm (s)}$ is protected from quantum corrections at large $N$. For the double-trace contribution, we apply the Lehmann--Symanzik--Zimmerman (LSZ) reduction formula to the four-point correlation function of scalar operators found in Refs.~\cite{Grabner:2017pgm,Gromov:2018hut} and
derive a closed-form expression for $A^{\rm (d)}$
that is valid  for any coupling $\xi^2$. We verify that at weak coupling it agrees with the result of an explicit
five-loop calculation. We study the properties of the double-trace amplitude $A^{\rm (d)}$ in the high-energy
limit and derive the exact expression for the leading Regge trajectory.

The paper is organized as follows. In the next section, we define single- and double-trace contribution to the four-particle scattering amplitude \re{color-ordered} in the conformal fishnet theory \re{L-dt}. In section~3, we present a five-loop calculation of the double-trace 
amplitude $A^{\rm (d)}$ in the large-$N$ limit.  In section~4, we apply the LSZ reduction procedure to the four-point correlation function of scalar operators and obtain an all-loop representation for $A^{\rm (d)}$. In section~5, we use this representation to 
examine the properties of $A^{\rm (d)}$ in the high-energy limit. Section~6 contains concluding remarks. Some details of the calculation
are summarized in three appendices.

\section{Four-particle amplitudes in  fishnet theory}

The four-particle amplitudes \re{color-ordered} 
can be classified according to the type   ($ X,\bar X, Z,\bar Z$) of scattered scalar particles. 
For the amplitude $\mathcal A$ to be nonzero, the total $U(1)\times U(1)$ charge should vanish. 
This leaves us with three nontrivial amplitudes: $\mathcal A_{XZ\bar X\bar Z}$, $\mathcal A_{XX\bar X\bar X}$ and $\mathcal A_{ZZ\bar Z\bar Z}$.
In addition, the invariance of \re{L-dt} under $X\to Z^{\rm t}$ and $Z\to  X^{\rm t}$ (with the conjugated fields transforming accordingly)
implies that the amplitudes $\mathcal A_{XX\bar X\bar X}$ and $\mathcal A_{ZZ\bar Z\bar Z}$ coincide up to an exchange of particles.

In general, scattering amplitudes in massless theories suffer from infrared (IR) divergences and require 
introducing a regulator, e.g. dimensional regularization with $D=4-2\epsilon$. We show below that the 
partial amplitudes $A^{\rm (s)}$ and $A^{\rm (d)}$ are IR-finite in the large-$N$ limit, and therefore they can be defined in $D=4$ dimensions.~\footnote{Infrared divergences do appear in $A^{\rm (s)}$ and $A^{\rm (d)}$ but at the level of the color suppressed corrections.}  As they are dimensionless functions of the Mandelstam invariants $s_{ij}=(p_i+p_j)^2$ and the coupling constant $\xi^2$ (we recall that the double-trace couplings are
given by \re{fixed} at the fixed point),  the partial amplitudes $A^{\rm (s)}$ and $A^{\rm (d)}$ have the following general form at large $N$~\footnote{As was shown in Ref.~\cite{Chicherin:2017bxc},  the amplitudes  \re{z} automatically satisfy the conformal Ward identities.}
\begin{align}\label{z}
A^{\rm (s,d)} = A^{\rm (s,d)}(z,\xi^2)\,,\qquad \qquad z= 1+ {2s_{13}\over s_{12}} 
\,.
\end{align}
The possible values of $z$ depend on the choice of scattering channel. In particular, for the process $1+2\to 3+4$ we have
$z=\cos\theta$, where $0\le\theta\le\pi$ is the scattering angle in the center-of-mass frame.
 
As was mentioned above, there are only two nontrivial four-particle amplitudes, $\mathcal A_{XZ\bar X\bar Z}$ and $\mathcal A_{XX\bar X\bar X}$. 
Let us first examine the former amplitude. In the Born approximation, $\mathcal A_{XZ\bar X\bar Z}$ 
receives contributions from the single-trace interaction term in \re{L0} and from the double-trace interaction terms in \re{L-dt} proportional to $\alpha_2^2$.
Replacing $\alpha_2^2$ by its value \re{fixed}, we find the corresponding  single- and double-trace partial amplitudes 
\begin{align}\label{Born-amp}
A^{\rm (s)}_{XZ\bar X\bar Z} = (4\pi\xi)^2\,,\qqqquad A^{\rm (d)}_{XZ\bar X\bar Z} = A^{\rm (d)}_{X\bar  Z\bar X Z} =-(4\pi\xi)^2\,.
\end{align}
All remaining partial amplitudes vanish. 

\begin{figure}[h!]
 \centering
 \psfrag{x1}[cc][cc]{$1$}\psfrag{x2}[cc][cc]{$2$}\psfrag{x3}[cc][cc]{$3$}\psfrag{x4}[cc][cc]{$4$}
 \includegraphics[width = 95mm]{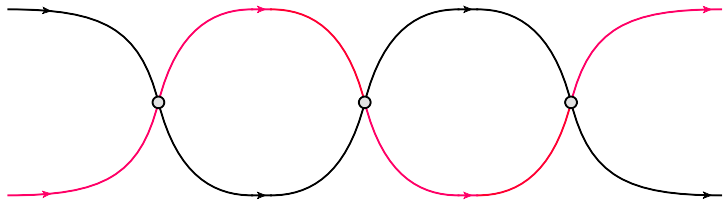}
\caption{Feynman diagrams contributing to the scattering amplitude $\mathcal A_{XZ\bar X\bar Z}$ in the large-$N$ limit. Incoming and outgoing black lines denote scalar particles $X$ and $\bar X$, respectively. In a similar fashion, red lines denote scalars $Z$ and $\bar Z$. 
Grey blob represents the sum of single- and double-trace vertices. } 
\label{fig:xz}
\end{figure}

Beyond the Born approximation, the leading-color contribution to $\mathcal A_{XZ\bar X\bar Z}$ comes from the diagrams shown in Fig.~\ref{fig:xz}.  
Each quartic vertex in these diagrams represents the sum of the single-trace $\xi^2-$interaction term  
and the double-trace $\alpha_2^2-$interaction terms defined in \re{L0} and \re{L-dt}, respectively. 
The contribution of 
the double-trace interaction terms proportional to $\alpha_1^2$ is suppressed by powers of $1/N$. 
Going through the color algebra, we find that, independently of the (single- or double-trace) type of vertices, the diagrams shown in Fig.~\ref{fig:xz}  
 produce a double-trace contribution.~\footnote{Naively one might expect that diagrams built from single-trace vertices could
produce a single-trace contribution. This does not happen due to the particular, chiral, form of the interaction term in  \re{L0}.} 
 Due to the additional minus
sign in front of $\alpha_2^2$ in \re{L-dt}, it is accompanied by powers of $(\xi^2-\alpha_2^2)$.  Since $\alpha_2^2=\xi^2$ at the 
fixed point \re{fixed}, the diagrams shown in Fig.~\ref{fig:xz}  vanish to all loops.
Thus,  $\mathcal A_{XZ\bar X\bar Z}$ is protected from loop corrections in the large-$N$ limit and the nonzero partial amplitudes are given
by  the Born level expressions \re{Born-amp}.

\begin{figure}[h!]
 \centering
 \psfrag{x1}[cc][cc]{$1$}\psfrag{x2}[cc][cc]{$2$}\psfrag{x3}[cc][cc]{$3$}\psfrag{x4}[cc][cc]{$4$}
 \includegraphics[width = 125mm]{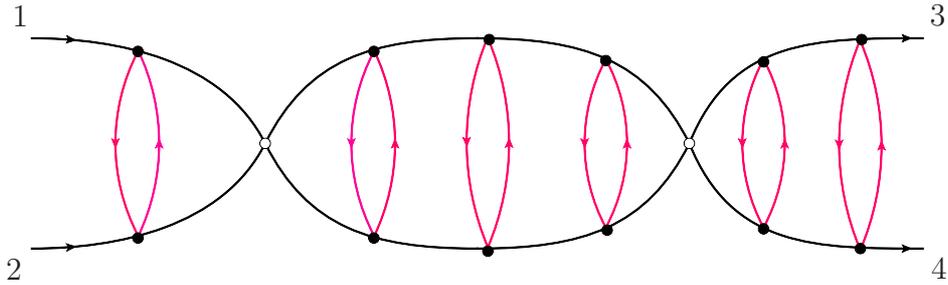}
\caption{Feynman diagrams contributing to the scattering amplitude $\mathcal A_{XX\bar X\bar X}$.
Black and white blobs represent the single- and double-trace vertices, respectively. } 
\label{fig:xx}
\end{figure}

Let us now examine the amplitude
$\mathcal A_{XX\bar X\bar X}$. The leading-color contribution to $\mathcal A_{XX\bar X\bar X}$ comes from the diagrams shown in Fig.~\ref{fig:xx}.
They
contain an arbitrary number of single-trace $\xi^2-$vertices and double-trace $\alpha_1^2-$vertices, as the contribution of $\alpha_2^2-$vertices is  
suppressed at large $N$. Each individual diagram  in Fig.~\ref{fig:xx} is IR finite but it contains UV divergent scalar loops. 
The UV divergences cancel however in the sum of all diagrams at the fixed point \re{fixed}. 

 In distinction to $\mathcal A_{XZ\bar X\bar Z}$, the amplitude
$\mathcal A_{XX\bar X\bar X}$ is not protected from quantum correction.  The diagrams shown in Fig.~\ref{fig:xx} produce a double-trace contribution to $\mathcal A_{XX\bar X\bar X}$  of the form
\begin{align}\label{A} 
\mathcal A_{XX\bar X\bar X}  = (2\pi)^4 \delta^{(4)}\Big(\sum_i p_i\Big)   \tr(T^{a_1} T^{a_2})  \tr(T^{a_3} T^{a_4}) \, A(z,\xi^2) \,,
\end{align}
where the particle with index $i$ carries the on-shell momentum $p_i$ and the color charge $T^{a_i}$. 
The double-trace partial amplitude $A(z,\xi^2)$ is a nontrivial UV- and IR-finite function of the coupling $\xi^2$ and the kinematical variable $z$ defined in \re{z}. 
The symmetry of $\mathcal A_{XX\bar X\bar X}$ under the exchange of particles $1$ and $2$ leads to an invariance of $A(z,\xi^2)$ under $z\to -z$.
Our goal in the rest of the paper is to find an expression for $A(z,\xi^2)$ at any coupling $\xi^2$. 

The vanishing of the quantum corrections to the single-trace ccmponent of $\mathcal A_{XZ\bar X\bar Z}$ and $\mathcal A_{XX\bar X\bar X}$ is in agreement with the properties of the four-particle amplitudes in planar $\mathcal N=4$ SYM. 
We recall that the latter amplitudes are given by the BDS ansatz and their dependence on the coupling is described by the cusp
anomalous dimension. The fishnet theory arises from $\mathcal N=4$ SYM in the double scaling limit described in the Introduction.
The cusp anomalous dimension vanishes in this limit and, as a consequence, the planar four-particle amplitudes cease depending on the coupling constant.
 
\section{Double-trace amplitude}

In this section, we compute the double-trace amplitude $A(z,\xi^2)$ at weak coupling. In the large-$N$ limit, 
$A(z,\xi^2)$ is given by the sum of Feynman diagrams shown in Fig.~\ref{fig:xx}.  As was mentioned in the previous section, each of these diagrams is UV divergent but the divergences cancel in their sum at the fixed point \re{fixed}. In what follows 
we employ a dimensional regularization with $D=4-2\epsilon$.   
 
\subsection{Scattering amplitude  at weak coupling}\label{sect:weak}

At tree level, the scattering amplitude \re{A} is generated by the double-trace $\tr(X^2)\tr(\bar X^2)$ interaction term  in \re{L-dt}
\begin{align}\label{Born}
A^{(0)} = 4 (4\pi\alpha_1)^2\,,
\end{align}
where the coupling $\alpha_1^2$ is given by \re{fixed}.

At one loop, the amplitude is given by the sum of two diagrams containing single- and double-trace vertices (shown by white and black blobs, respectively)
\begin{align}\label{1loop} 
A^{(1)} {}&= \parbox[c]{35mm}{\psfrag{+}[cc][cc]{$+$} \includegraphics[width = 35mm]{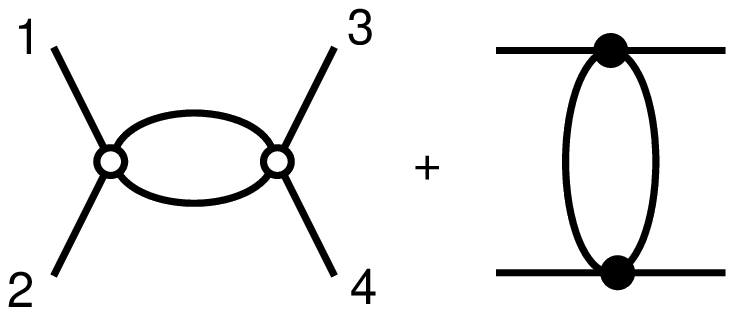}}
= 4(4\pi \alpha_1)^4 \pi(s_{12}) + (4\pi \xi)^4\pi(s_{13}) + (p_1\leftrightarrow p_2)\,.
\end{align}
Here $\pi(s)$ denotes the one-loop scalar integral 
\begin{align}\label{pi1}
\pi(s) = \int {d^{4-2\epsilon} \ell \over i \, (2\pi)^{4-2\epsilon}} {1\over \ell^2 (K-\ell)^2} ={(-s/\mu^2)^{-\epsilon}\over (4\pi)^{2-\epsilon}}
{\Gamma(\epsilon)\Gamma^2(1-\epsilon)\over \Gamma(2-2\epsilon)}  \,,
\end{align}
where $s=K^2$ and $\mu^2$ is a UV cut-off. Both diagrams in \re{1loop} develop a $1/\epsilon$ UV pole. We verify that at the fixed point \re{fixed}, for $\alpha_1^2=\alpha_+^2$, the poles cancel  in their sum leading to
\begin{align} 
A^{(1)} {}&= - (4\pi)^2 \xi^4 \ln {z-1\over 2} + (z\to -z)\,,
\end{align}
where $z$ is defined in \re{z}.

At two loops, we have 
\begin{align} \label{2loop}
A^{(2)} {}&=\parbox[c]{40mm}{ \psfrag{+}[cc][cc]{$+$}\includegraphics[width = 40mm]{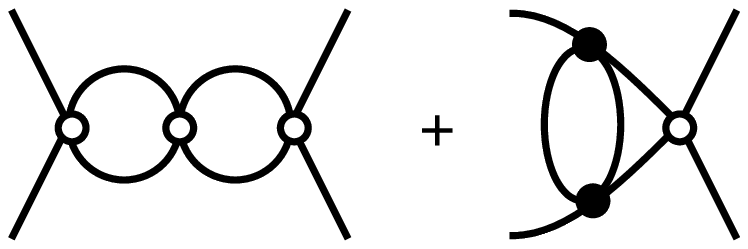}}
  = 8(4\pi \alpha_1)^6 \left[\pi(s_{12})\right]^2 +  4(4\pi)^6 \alpha_1^2\,\xi^4 V(s_{12}) + (p_1\leftrightarrow p_2)\,.
\end{align}
Here the first diagram factorizes into the product of one-loop scalar integrals \re{pi1},  the second one involves the  vertex integral
$V(s_{12})$ 
\begin{align} \label{V}\notag
V(s_{12}) 
{}& =   \int {d^{4-2\epsilon} \ell \over i (2\pi)^{4-2\epsilon}} {\pi(\ell^2) \over (\ell+p_1)^2 (\ell-p_2)^2}  
\\[2mm]
{}&
=   {(-s_{12}/\mu^2)^{-2\epsilon}\over (4\pi)^{4-2\epsilon}}
{\Gamma(\epsilon)\Gamma(2\epsilon)\Gamma^2(1-\epsilon)\Gamma^2(1-2\epsilon) \over \Gamma(2-2\epsilon)\Gamma(2-3\epsilon)}\,.
\end{align}
For arbitrary $\alpha_1^2$ the expression on the right-hand side of \re{2loop} develops a double UV pole $1/\epsilon^2$.
Combining together \re{Born}, \re{1loop} and \re{2loop} and replacing the double-trace coupling by its value at  the fixed 
point~\footnote{To get an analogous expression at the second fixed point $\alpha_1^2=\alpha_-^2$ it suffices to replace $\xi^2\to -\xi^2$.}, 
$\alpha_1^2=\alpha_+^2$, we find that all the UV poles cancel leading to the following expression for the two-loop amplitude
\begin{align}\label{A-2loops}
A  =  i (4\pi \xi)^2\left[1+ i \xi ^2 \left(\ln  \frac{z-1}{2} +1\right)+\xi ^4   \left(\frac32+ \frac{\pi
   ^2}3\right) +O\left(\xi ^6\right)\right] + (z\to -z)\,.
\end{align}
In agreement with our expectations, it is free from any divergences. 

We notice that the two-loop correction to the amplitude \re{A-2loops} does not depend on $z$. As we show in a moment, the same pattern persists at every even loop order, so that 
the nontrivial dependence of the amplitude on the kinematical invariants only comes from odd loops. To make this property explicit,
it is convenient to separate $A(z,\xi^2) $ into the sum of even and odd functions of $\xi^2$
\begin{align}\label{A-plus}
A(z,\xi^2) = A_+ (z,\xi^2) +  A_-(\xi^2) \,,
\end{align} 
where $A_+(z,-\xi^2)=A_+(z,\xi^2)$ and $A_-(-\xi^2)= -A_-(\xi^2)$. At weak coupling $A_-(\xi^2)$ receives contributions from even loops, and is expected to
be $z-$independent.

To show this, we examine the Feynman diagrams contributing to $A$ at three and four loops: 
\begin{align}\label{3loop}
A^{(3)}{}& =\parbox[c]{95mm}{\psfrag{+}[cc][cc]{$+$} \includegraphics[width = 95mm]{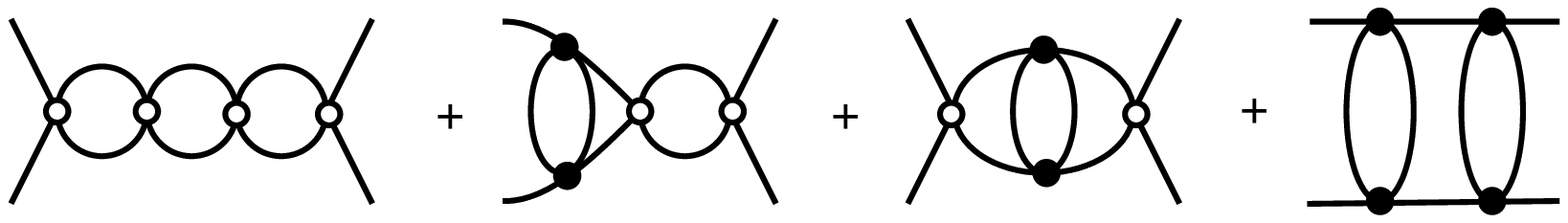}}
\\[4mm]\label{4loop}
A^{(4)}{}& =\parbox[c]{135mm}{\psfrag{+}[cc][cc]{$+$} \includegraphics[width = 135mm]{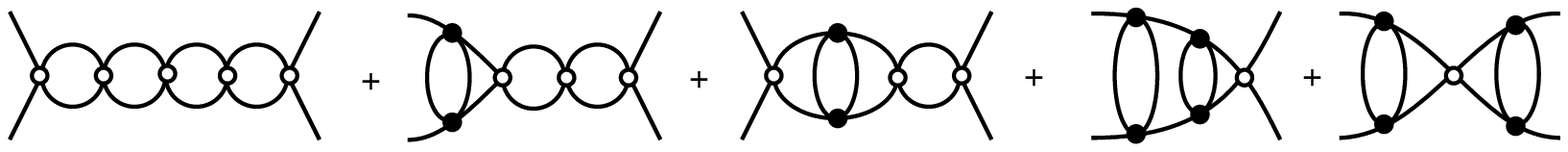}}
\end{align}
\medskip

\noindent
It is easy to see that, similar to \re{2loop},  all diagrams on the right-hand side of \re{3loop} and 
\re{4loop}, except the right-most diagram in $A^{(3)}$, factor out into the product of two- 
and three-point functions.
They develop UV divergences and 
depend on $s_{12}/\mu^2$ (but not on $s_{13}/\mu^2$). The right-most diagram in \re{3loop} also produces a UV divergence, but it depends on two dimensionless ratios, $s_{12}/\mu^2$ and $s_{13}/s_{12}$.

Because the amplitude is UV finite, the dependence on $\mu^2$ should disappear in the sum of all diagrams. 
At four loops, the contributing  diagrams only depend on $s_{12}/\mu^2$. Being $\mu^2-$independent, their sum ought to be a constant. At three loops, the additional $z-$dependence arises from the right-most ladder-like diagram in \re{3loop}. 
Going to higher loops, we find that the $z-$dependent contribution can only come from analogous ladder-like diagrams.
Such diagrams
are built from even numbers of single-trace vertices and, therefore, 
can only appear at odd loops. This explains why the
$z-$dependent contribution to the scattering amplitude \re{A-plus} is accompanied by {even} powers of $\xi^2$. 

As follows from the above analysis, the $\xi^2-$odd part of the amplitude, $A_-(\xi^2)$, does not depend on the kinematical invariants and  is given by a sum of factorizable diagrams.  
At the same time, the $\xi^2-$even part of the amplitude takes the following general form 
\begin{align}\label{diags}
A_+ (z,\xi^2)=\left( \parbox[c]{85mm}{\psfrag{+}[cc][cc]{$+$}\psfrag{dots}[cc][cc]{$\ldots$} \includegraphics[width = 85mm]{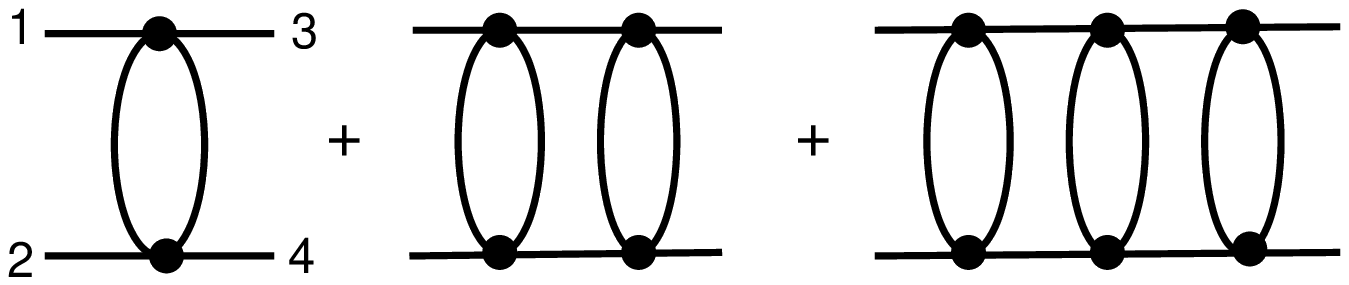}}\right) - \text{[UV div]},  
\end{align}
where the dots denote higher-order ladder diagrams as well as diagrams with the legs $3$ and $4$ exchanged.  The last term on the right-hand side denotes the contribution of factorizable diagrams containing double-trace vertices. 
It is needed to restore
the UV finiteness of $A_+(z,\xi^2)$.

The evaluation of most of the Feynman diagrams in \re{3loop} and \re{4loop} is  straightforward. The ladder diagram can be computed using the Mellin-Barnes representation (see e.g. Ref.~\cite{Smirnov:2012gma}).
Going through the calculation, we find,~\footnote{Here we also included the five-loop contribution to $A_+$.}
\begin{align} \label{B-loops}\notag
A_-  {}&=  32i\pi^2\left[\xi^2 + \xi^6 \left(\frac32+ \frac{\pi
   ^2}3\right)  + \xi^{10}\left(-\frac{49}{8}+\frac{\pi
   ^2}{6}+\frac{2 \pi ^4}{45}\right) +O\left(\xi^{14}\right)\right] \,,
 \\[2mm] 
 A_+  {}&= 16\pi^2 \left[ \xi^4 f_1(z) + \xi^8 f_3(z)    + \xi^{12} f_5(z)      +O\left(\xi^{16}\right)\right] + (z\to -z)\,,
\end{align}
where $f_\ell(z)$ are nontrivial functions of the ratio of kinematical invariants \re{z} at $\ell$ loops. They admit a compact representation 
\begin{align}\notag\label{fun}
f_1{}=&- H_0 -1\,,
\\\notag
f_3{}=&-H_{-1, 0, 0}  - {\pi^2\over 2}  H_{-1} -4 \zeta_3  +3\,,
\\ \notag
f_5 {}=&-2 \zeta_3 H_{0,-1}+2 \zeta_3 H_{-1}-\pi ^2 H_{0,-1,-1}  +\pi ^2 H_{-1,-1}+\frac{\pi ^2}{3}   H_{0,-1,0}
 -\frac{\pi ^2}{3}   H_{-1,0}
 \\
{}& -2
   H_{0,-1,-1,0,0}+2H_{-1,-1,0,0}+14 \zeta_5-\frac{2 \pi ^2
   \zeta_3}{3}+\frac{2 \pi ^2}{3}-12\,,
\end{align}
where $\zeta_n$ are Riemann zeta values and $H_{a_1,a_2,\dots} \equiv H_{a_1,a_2,\dots}((z-1)/2)$ are harmonic polylogarithms \cite{Remiddi:1999ew,Maitre:2005uu}.
In general, $f_\ell(z)$ is given by a linear combination of  functions $H_{a_1,\dots,a_w}$ which carry indices $a_i\in\{0,-1\}$ whose total number $w$, or equivalently the weight, satisfies  $w \le \ell$.
In the next section, we derive the exact expressions for $A_\pm$ that are valid for any coupling.

We would like to emphasize that the relations \re{B-loops} hold at the fixed point $\alpha_1^2=\alpha_+^2$. At the second fixed point, for $\alpha_1^2=\alpha_-^2$, the 
scattering amplitude is given by
$A(z,-\xi^2) = A_+ (z,\xi^2) -  A_-(\xi^2)$.

\subsection{High-energy limit at weak coupling}\label{sect:high}

The explicit expressions for the functions \re{fun} become rather lengthy at high loops. It is instructive to examine \re{fun} in two limiting cases:
 $z=1$ and $z\to\infty$. According to \re{z}, they correspond to two different high-energy limits,  
\begin{align}
s_{12} \gg s_{13}  \,,\qqqquad  s_{13} \gg s_{12} \,,
\end{align}
respectively. Applying the Regge theory \cite{Gribov:2003nw}, we expect that in both cases  the asymptotic behaviour of the 
scattering amplitude is governed by Regge trajectories exchanged in the $t-$channel of the corresponding processes $1+2\to 3+4$ and $1+3\to 2+4$.  
 
For $s_{12} \gg s_{13}$, or equivalently $z \to 1$, we find from \re{A-plus} and \re{B-loops} that the amplitude has the following asymptotic behaviour
\begin{align}\label{A-regge1}
A(z,\xi^2) \stackrel{z\to 1}{=} 16\pi^2 \left[ \xi^4 \ln y + c(\xi^2) + O(y)  \right],
\end{align}
where $y=(z-1)/2=s_{13}/s_{12}$ and $c(\xi^2)$ is a series in $\xi^2$ with constant coefficients whose explicit form is not important for us. Notice that the asymptotic behavior  $A\sim \ln y$ is one-loop 
exact,  while high-order corrections only contribute to the constant part.

For $s_{13} \gg s_{12}$, or equivalently $z\to\infty$, we find from \re{fun} that, in distinction to the previous case, the higher order corrections to the 
amplitude are enhanced by powers of $\ln (z/2)\sim \ln (s_{13}/s_{12})$,
\begin{align}\notag\label{A-regge2}
{}&f_1= -\ln (z/2)-1+\dots\,,
\\\notag
{}&f_3= -\frac{1}{6} \ln ^3 (z/2)  - {\pi ^2\over 2} \ln (z/2) + \dots\,,
\\
{}& f_5=-\frac{1}{60} \ln ^5(z/2)+\frac{1}{12}\ln
   ^4(z/2) -\frac{1}{9} \pi ^2 \ln ^3(z/2)+\dots \,,
\end{align}
where the dots denote subleading terms.

The relations \re{A-regge1} and \re{A-regge2} are in agreement with the Regge theory expectations \cite{Gribov:2003nw}. In the high-energy limit, the leading  contribution to the amplitude \re{diags} comes from diagrams with the minimal number of scalars exchanged. For the process $1+2\to 3+4$, we find from \re{diags} that the number of particles exchanged in the $s_{13}-$channel increases with the loop order and, therefore, the dominant contribution only comes from the one-loop diagram leading to \re{A-regge1}. 

For the process $1+3\to 2+4$, the ladder diagrams in \re{diags} describe the propagation of two scalars in the $s_{12}-$channel, interacting through 
the exchange of  a pair of scalars (see \re{unit} below). In the leading logarithmic approximation (LLA), $L=\xi^2 \ln (z/2)={\rm fixed}$ for $z\to\infty$, the dominant 
contribution to the amplitude comes from the integration over the loop momenta in the multi-Regge kinematics (corresponding to the  strong ordering of rapidity of the
exchanged scalars, see \re{Sudakov} and \re{order}). Going through the  calculation we get (see Appendix~\ref{app:C} for details)
\begin{align}\label{f-LLA}
f_{2n+1}=  -{ \ln^{2n+1} (z/2)\over (2n+1)n! (n+1)!} +\dots \,.
\end{align}
We verify that this relation correctly reproduces the first term on the right-hand side of \re{A-regge2}.

Substitution of \re{f-LLA} into \re{B-loops} yields the following result for the amplitude 
\begin{align}\notag\label{A-lla}
A_{\rm LLA} {}&= - (4\pi \xi)^2 \sum_{n\ge 1} {L^{2n+1} \over (2n+1)n! (n+1)!}  + (z\to -z)
\\
{}&= -(4\pi\xi)^2 \dashint_{-1}^1 {dx \over \pi x}\sqrt{1-x^2}  \e^{2Lx} + (z\to -z)\,,
\end{align}
where $L=\xi^2 \ln (z/2)$ and the integral is defined using the principal value prescription~\footnote{The sum in the first relation of \re{A-lla} can be expressed in terms of Bessel and modified Struve functions. \label{foot}}.
For $L\gg 1$, the dominant contribution comes from integration in the vicinity of $x=1$ 
 \begin{align}\label{A-LLA}
A_{\rm LLA} {}&=  -(4\pi\xi)^2{\e^{2L}\over 4\sqrt{\pi}}L^{-3/2}\left[1+\frac{9}{16} L^{-1}+\frac{345}{512}
   L^{-2}+\dots\right]+ (z\to -z)\,.
\end{align}
We deduce from this relation that, in the leading logarithmic approximation, the amplitude has the typical Regge behavior,
\begin{align}\label{A-regge}
A_{\rm LLA} \sim L^{-3/2}\ \e^{2L} \sim {z^{2\xi^2}\over (\ln z)^{3/2}}\,,
\end{align}
with $L=\xi^2 \ln (z/2)$ and $z\sim 2s_{13}/s_{12}$.
The presence of the factor $L^{-3/2}$ on the right-hand side implies that the corresponding Regge singularity is a cut rather than a pole. 

We recall that, in the expression for the scattering amplitude \re{A}, $A_{\rm LLA}(z,\xi^2)$ is accompanied by the double-trace color tensor
$\tr(T^{a_1} T^{a_2}) \tr(T^{a_3} T^{a_4})$ which projects the two pairs of particles onto a color-singlet state.
As a consequence, the Regge singularity exchanged in the $s_{12}-$channel carries zero color charge. The exponent of $z$ in \re{A-regge} defines the position of this singularity in the leading logarithmic approximation  
$ 
J_R=2\xi^2 + \dots\,,
$ 
where the dots denote subleading corrections. We derive the exact expression for $J_R$ in Sect.~\ref{sect:ex}. 

\section{Amplitude from correlation function}

The Feynman diagrams shown in Fig.~\ref{fig:xx}  have a nice iterative structure suggesting that they can be evaluated using
the Bethe-Salpeter approach. This turns out to be a nontrivial task because the external and internal lines in these diagrams are on-shell and off-shell,
respectively, and, therefore, cannot be treated on an equal footing.

In this section, we present another approach to computing the scattering amplitude \re{A}.  It relies on 
applying the Lehmann--Symanzik--Zimmerman (LSZ) reduction formula to the
four-point correlation function
\begin{align}\label{G}
G(x_1,x_2|x_3,x_4) = {1\over N^2}\vev{\tr(X(x_1) X(x_2))\tr(\bar X(x_3) \bar X(x_4))}\,,
\end{align}
which has been computed in  Refs.~\cite{Grabner:2017pgm,Gromov:2018hut}. 
Here the color indices of the scalar fields are contracted in such a way as to project the two pairs of scalars onto 
a color-singlet state and, thus, match the properties of the scattering amplitude \re{A}.
In the planar limit, the correlation function \re{G} is given by the sum of diagrams shown in Fig.~\ref{fig:xx} (defined in configuration space and the end-points $i$ having the coordinates $x_i$).
The main advantage of the correlation function is that all lines are off-shell and, therefore, can be treated in the same manner.

Following the LSZ procedure, we have to Fourier transform $G(x_1,x_2|x_3,x_4)$ and identify
the residue at the four simultaneous massless poles $p_i^2=0$
\begin{align}\label{LSZ}
\int \prod_i d^4 x_i \e^{i p_i x_i} G(x_1,x_2|x_3,x_4) = {1\over  p_1^2 p_2^2 p_3^2 p_4^2} \times (2\pi)^4 \delta^{(4)}(\sum_i p_i) A(z,\xi^2) + \dots\,,
\end{align}
where the dots denote terms subleading for $p_i^2\to 0$. 

\subsection{Exact correlation function}

Applying the Bethe-Salpeter approach and
using the conformal symmetry, we can obtain the following representation for  the correlation function \re{G}  (see Refs.~\cite{Grabner:2017pgm,Gromov:2018hut}), 
\begin{align}\label{G-OPE}
G(x_1,x_2|x_3,x_4) = \sum_{J\ge 0} \int_{-\infty}^\infty  {d\nu }  
{\mu(\nu,J)\over h(\nu,J)-\xi^4}
\Pi_{\nu,J}(x_1,x_2|x_3,x_4) 
+ (x_1\leftrightarrow x_2)\,,
\end{align}
where the sum runs over all states propagating in the OPE channel $x_{12}^2\equiv (x_1-x_2)^2 \to 0$. These states carry  Lorentz spin $J$ and  scaling 
dimension $\Delta=2+2i\nu$. Their contribution to \re{G-OPE} is described by the function,
\begin{align}\label{Pi}
\Pi_{\nu,J} =  \int d^4 x_0 \, \Phi^{\mu_1\dots\mu_J}_{\nu}(x_{10},x_{20})  \Phi^{\mu_1\dots\mu_J}_{-\nu}(x_{30},x_{40}) \,,
\end{align}
which is built out of the completely symmetric traceless tensors $\Phi^{\mu_1\dots\mu_J}_{\nu}$. They take the form of three-point spinning correlation functions, e.g.
\begin{align}\label{Phi}\notag
\Phi_{\nu,J} (x_{10},x_{20}) {}&= n_{\mu_1}\dots n_{\mu_J}\Phi^{\mu_1\dots\mu_J}_{\nu}(x_{10},x_{20})
\\
{}&={1\over x_{12}^2} \lr{x_{12}^2\over x_{10}^2 x_{20}^2}^{(\Delta-J)/ 2}\lr{{2(nx_{10})\over x_{10}^2}-{2(nx_{20})\over x_{20}^2}}^J,
\end{align}
with $x_{ij}\equiv x_i-x_j$ and $n^\mu$  an auxiliary light-like vector $(n^2=0)$. 

The functions \re{Phi}  belong to
the principal series of the conformal group. They form a complete orthogonal set of states, and the kinematical factor $\mu(\nu,J)$ defines their norm \cite{Dobrev:1977qv}
\begin{align}\label{c1}
\mu(\nu,J)={ \nu^2 (4\nu^2+(J+1)^2)(J+1) \over 2^{J+4}\pi^7}\,.
\end{align}
The dependence of the correlation function \re{G-OPE} on the coupling constant arises through the factor $1/(h(\nu,J)-\xi^4)$.
Due to the iterative structure of the Feynman diagrams  in Fig.~\ref{fig:xx}, it has the form 
of a geometric series  in $\xi^4/h(\nu,J)$ with the function $h(\nu,J)$ being the eigenvalue of the `graph generating kernel' entering the Bethe-Salpeter equation  (see Refs.~\cite{Grabner:2017pgm,Gromov:2018hut})
\begin{align}\label{Omega} 
h(\nu,J) = (\nu^2+J^2/4)(\nu^2+(J+2)^2/4)\,.
\end{align}

The relation \re{G-OPE} can be used to decompose  the correlation function over the conformal partial waves  in the OPE channel $ x_{12}^2\to 0$. 
Closing the integration contour  over $\nu$  in \re{G-OPE} into the lower half-plane and
picking up the residues at the poles located at $ h(\nu,J)=\xi^4$ we find
\begin{align}\label{G-cont}
G(x_1,x_2|x_3,x_4) = {1\over (x_{12}^2 x_{34}^2)^2} \sum_{J\ge 0} \, \sum_{\Delta=\Delta_2,\Delta_4}C_{\Delta,J} \, g_{\Delta,J}(u,v)\,.
\end{align}
Here $C_{\Delta,J}$ are the OPE coefficients squared and $g_{\Delta,J}(u,v)$ are the well-known four-dimensional conformal blocks depending on two cross-ratios, $u=x_{12}^2x_{34}^2/(x_{13}^2x_{24}^2)$ and 
$v=x_{23}^2x_{14}^2/(x_{13}^2x_{24}^2)$. 

For each Lorentz spin $J$, the sum in \re{G-cont} contains
the contribution of two primary operators. Their scaling dimensions $\Delta_2(J)$ and $\Delta_4(J)$ satisfy the relation
\begin{align}\label{Delta}
\Delta=2+ 2i\nu\, \Big|_{h(\nu,J)=\xi^4}\,,
\end{align}
subject to the additional condition $\Re \Delta >2$. At weak coupling, there are two solutions, $\Delta_2=2+J +O(\xi^2)$ and $\Delta_4=4+J +O(\xi^4)$,  
describing the operators with twist $2$ and $4$, respectively. The explicit expression for the scaling dimensions $\Delta_{2,4}$ and the OPE coefficients $C_{\Delta,J}$ can be found in Ref.~\cite{Gromov:2018hut}.

\subsection{LSZ reduction}

To obtain the scattering amplitude, we have to substitute  \re{G-OPE} into \re{LSZ}, perform a Fourier transform with respect 
to the external points and identify the residue at the four massless poles. Due to the factorized form of \re{Pi}, this amounts to finding the residue of the functions \re{Phi} on the two-particle pole
\begin{align}\label{Phi-F1}
\lim_{p_i^2\to 0} p_1^2\, p_2^2 \int d^4 x_1 d^4 x_2 \e^{ip_1 x_1+ip_2 x_2} \Phi_{\nu}^{\mu_1\dots\mu_J}(x_{10},x_{20}) =  (2\pi)^4
{\e^{ix_0(p_1+p_2)} }\,\widetilde \Phi_{\nu}^{\mu_1\dots\mu_J}(p_1,p_2) \,,
\end{align}
where $\widetilde \Phi_{\nu}^{\mu_1\dots\mu_J}(p_1,p_2)$ is a completely symmetric traceless tensor depending on the light-like vectors $p_1$ and $p_2$.
The function $\widetilde \Phi_{-\nu}^{\mu_1\dots\mu_J}(p_3,p_4)$ is defined in a similar manner. 

Then, we apply \re{Phi-F1} to obtain from \re{LSZ} and \re{G-OPE} the following representation for the scattering amplitude
\begin{align}\label{amp-0}
A(z,\xi^2)=(2\pi)^8\sum_{J\ge 0} \int_{-\infty}^\infty  {d\nu }  
{\mu(\nu,J)\over h(\nu,J)-\xi^4} \Omega_{\nu,J}(z)+ (z\to -z)\,.
\end{align}
Here $\Omega_{\nu,J}(z)$  is given by the product of two tensors \re{Phi-F1} with all Lorentz indices
contracted
\begin{align}\label{P-def}
\Omega_{\nu,J}(z) =  \widetilde \Phi_{\nu}^{\mu_1\dots\mu_J}(p_1,p_2)\,\widetilde \Phi^{\mu_1\dots\mu_J}_{-\nu}(p_3,p_4)\,.
\end{align}
It depends on the ratio of kinematical invariants \re{z} as well as on the quantum numbers of the exchanged states.

As before, it is convenient to project the Lorentz indices on both sides of \re{Phi-F1} onto an auxiliary light-like vector and define $ \widetilde \Phi_{\nu,J}\equiv n_{\mu_1}\dots n_{\mu_J}\widetilde \Phi_{\nu}^{\mu_1\dots\mu_J}(p_1,p_2) $. Substituting \re{Phi} into  \re{Phi-F1} and
going through the calculation we find  
(see Appendix~\ref{app:A}),
\begin{align}\label{Phi-F}
 \widetilde \Phi_{\nu,J} (p_1,p_2) {}&
 = i^J   {(-s_{12}/4)^{{t}/{2}-1} \Gamma(t-1) \Gamma(J+1)\over \Gamma(t/2)\Gamma(J+t/2)\Gamma(J+t-1)}\, (\xi_1+\xi_2)^J C_J^{(t-1)/2}
 \lr{\xi_1-\xi_2\over \xi_1+\xi_2}\,,
\end{align}
where $t=\Delta-J=2+2i\nu-J$  and   $s_{12}=(p_1+p_2)^2$. Here 
$\xi_i=(p_i n)$, and  $C_J^{(t-1)/2}$ is a  Gegenbauer polynomial.

We can recover $\widetilde \Phi_{\nu}^{\mu_1\dots\mu_J}(p_1,p_2)$ by acting on \re{Phi-F} with  the differential operators \cite{Dobrev:1975ru},
\begin{align}\label{D-op}
\widetilde \Phi_{\nu}^{\mu_1\dots\mu_J}(p_1,p_2) = {1\over (J!)^2} {\mathcal D}^{\mu_1} \dots {\mathcal D}^{\mu_J}\widetilde \Phi_{\nu,J} (p_1,p_2)\,,
\end{align}
where ${\mathcal D}^\mu=(1+ (n\partial_n)) \partial_n^\mu - n^\mu \partial_n^2/2$ and $\partial_n^\mu=\partial/\partial n_\mu$.
Substituting \re{D-op} into \re{P-def} we find that $\Omega_{\nu,J}$ is a dimensionless scalar function of the ratio of Mandelstam  invariants $s_{13}/s_{12}$. 

According to \re{P-def},
the dependence of 
 $\Omega_{\nu,J}$ on $s_{13}$ can only arise from the contraction of Lorentz indices in the product of the two tensors.
 Since the number of indices matches the Lorentz spin,
$\Omega_{\nu,J}$ ought to be a polynomial of degree $J$ in $s_{13}$ or equivalently in $z$. In addition, it has the parity properties
 \begin{align}\label{par}
\Omega_{\nu,J} (z)=(-1)^J \Omega_{\nu,J} (-z)\,,\qqqquad
\Omega_{-\nu,J} (z)=\Omega_{\nu,J} (z)     \,.
\end{align}
Indeed, it follows from \re{Phi} and \re{Phi-F1} that
 $\widetilde \Phi_{\nu}^{\mu_1\dots\mu_J}(p_1,p_2)$ acquires the sign factor $(-1)^J$ under the exchange of the momenta $p_1$ and $p_2$. The same transformation acts as $z\to -z$  leading to the first relation in \re{par}.
The second relation in \re{par} follows from the invariance of both sides of \re{P-def} under the exchange of momenta, $p_1\leftrightarrow p_3$ and $p_2\leftrightarrow p_4$.  Combining these properties together, we conclude that $\Omega_{\nu,J}(z) $ has the following general form, 
\begin{align}\label{Om1}
\Omega_{\nu,J}(z) = Q_{\nu,J}z^J + O(z^{J-2})\,,
\end{align}
where $Q_{\nu,J}$ and all the subleading expansion coefficients are even functions of $\nu$.

The explicit expression for the polynomial \re{Om1} can be found from \re{P-def} and \re{Phi-F} (details of the calculation are given in 
Appendix~\ref{app:B})
\begin{align}\label{P-grow}
\Omega_{\nu,J}(z) =\frac{2^J}{\pi ^2
   } \sinh ^2\left( \pi  \nu+ i {\pi  J}/{2} \right) \sum_{k=0}^J \frac{P_k(z) P_{J-k}(z) }{ 
    \left({J}/{2}-k\right)^2+\nu ^2 }  \,,
\end{align}
where $P_J(z)$ is a Legendre polynomial. It also admits a representation as a linear combination of Legendre polynomials,
\begin{align}\notag\label{Om2}
\Omega_{\nu,J}(z) = (-2)^J  { \sinh (2 \pi  \nu )\over 4 \pi ^2 \nu}
\sum_{k=0}^{[J/2]}
{}&
\frac{
 (2 J-4k+1) \Gamma \left(k+\frac{1}{2}\right)
   \Gamma (J-k+1) }{  \Gamma (k+1) \Gamma
   \left(J-k+\frac{3}{2}\right)  }
 \\
   \times {}& \frac{  \Gamma \left(\frac{J}2-k- i \nu +\frac12\right)
   \Gamma \left(\frac{J}2-k+ i \nu +\frac12 \right)}{ \Gamma \left(\frac{J}{2}-k-i \nu +1\right) \Gamma
   \left(\frac{J}{2}-k+i \nu +1\right)}
  { P}_{J-2 k}(z) \,,
\end{align}
where $[J/2]$ stands for the entire part.

For even/odd spin $J$ the sum in \re{Om2} contains Legendre polynomials with even/odd indices. It is easy to verify that the relations \re{P-grow} and \re{Om2} satisfy \re{par}. At large $z$ we match \re{Om2} into \re{Om1} to find
\begin{align} 
\label{Q-coef}
Q_{\nu,J}  
   {}&= (-1)^J \frac{\sinh (2 \pi  \nu )}{2 \pi  \nu } \frac{\Gamma (J-2 i \nu +1) \Gamma (J+2 i \nu +1)}{ \left[\Gamma \left(\frac{J}{2}-i \nu +1\right) \Gamma \left(\frac{J}{2}+i \nu
   +1\right)\right]^2}\,.
\end{align} 

Finally, we substitute \re{c1} and \re{Omega} into \re{amp-0} and obtain the following representation for the scattering amplitude
\begin{align}\label{A4-int} 
A(z,\xi^2)  {}& =  8\pi \int_{-\infty}^\infty d\nu \sum_{J \ge 0} \,  \frac{2^{1-J} (J+1) \left((J+1)^2+4 \nu ^2\right)\nu ^2 }{
   \left( \nu ^2+ {J^2}/{4}\right) \left(\nu ^2+ (J+2)^2/4\right)-\xi
   ^4 } \, \Omega_{\nu,J} (z) +(z\to -z)\,,
\end{align}
which is valid for arbitrary coupling $\xi^2$.
Here the last term on the right-hand side is needed to restore the crossing symmetry of the amplitude. By virtue of \re{par},  this amounts
to retaining the contribution of even spins $J$ only.
In the next section, we apply \re{A4-int}  to compute the amplitude in the high-energy limit.
 
The following comments are in order. 

It is not obvious \textit{a priori} that the $\nu-$integral in \re{A4-int} is convergent. For real $\xi^2$, the integrand 
in \re{A4-int} contains poles on the real $\nu-$axis. For the $\nu-$integral to be well-defined, the coupling constant should have a nonzero imaginary part. 
The same property has been previously observed in Ref.~\cite{Gromov:2018hut} for the correlation function \re{G-OPE}. It reflects the fact that,  
 as functions of the coupling constant, 
the various quantities in the conformal fishnet theory (correlation functions, scattering amplitudes) 
have a branch cut  for positive $\xi^4$.

In addition, it follows from \re{P-grow} that $\Omega_{\nu,J}(z)\sim \sinh^2(\pi\nu)/\nu^2$ at large $\nu$ and, therefore, the $\nu-$integral  
in \re{A4-int} diverges at infinity for any given $J$. A close examination shows however 
that the divergences cancel in the sum over 
all spins. This property can be made manifest by applying the Watson-Sommerfeld transformation to \re{A4-int}, 
\begin{align}\label{WS}
A (z,\xi^2){}& =   \int_C {dJ\over 2\pi i} { 8\pi^2\over\sin(\pi J)} \int_{-\infty}^\infty d\nu   \frac{2^{1-J} (J+1) \left((J+1)^2+4 \nu ^2\right)\nu ^2 }{
   \left( \nu ^2+ {J^2}/{4}\right) \left(\nu ^2+ (J+2)^2/4\right)-\xi
   ^4 } \, \Omega_{\nu,J} (z) +(z\to -z)\,,
\end{align}
where the integration contour $C$ encircles the nonnegative integer $J$ in an anti-clockwise direction. Exchanging the order of integrations in \re{WS} we verify using  \re{Cheb} that the divergences at large $\nu$ are accompanied by the  vanishing integrals of the form $ \int_C  (dJ/\sin(\pi J)  ) (J+1)^{1+2k} U_J(z) =0$ (with $k$ nonnegative integer).

The relation \re{WS} is the main result of this paper. The representation
\re{WS} has a striking similarity with the analogous expression for the high-energy asymptotics of 
the off-shell amplitudes in the conformal Regge theory \cite{Costa:2012cb}. In distinction to the latter, the relation \re{WS} holds 
for on-shell amplitudes and in arbitrary kinematics. As we show in the next section, in the high-energy limit, for $z\to \infty$, the 
asymptotic behavior of the amplitude
\re{WS} is governed by the Regge singularities of the integrand of \re{WS} in the complex $J-$plane. In this limit, the relation \re{WS}
agrees with the results of Ref.~\cite{Costa:2012cb}.

\subsection{Integral over $\nu$}

The integrand in  \re{A4-int} has poles in the $\nu-$plane located at
\begin{align}\label{h-eq}
h(\nu,J) =\left(\nu^2+{J^2\over 4}\right)\left(\nu^2+{(J+2)^2\over 4}\right)= \xi^4\,.
\end{align}
This suggests to deform the integration contour in \re{A4-int} and evaluate the $\nu-$integral by residues. 
Among the four solutions to \re{h-eq}, two are located in the lower half-plane.
At weak coupling they are given by
\begin{align}\notag\label{nus}
{}&\nu_2(J)=-\frac{i J}{2}+\frac{i \xi ^4}{J (J+1)}+O\left(\xi ^8\right)\,,
\\
{}&\nu_4(J)=-\frac{i}{2} (J+2)-\frac{i \xi ^4}{(J+1) (J+2)}+O\left(\xi ^8\right)\,,
\end{align}
where the expansion runs in powers of $\xi^4$.
We recall that the solutions to \re{h-eq} also define the scaling dimensions \re{Delta} of the operators that contribute to the four-point correlation
function \re{G-cont}. The subscript in $\nu_2(J)$ and $\nu_4(J)$ refers to the twist of the exchanged operators. 

Notice that the first relation in \re{nus} is not well-defined for $J=0$, The reason for this is that $\nu\sim \sqrt{\xi^4-J^2/4}$ at small $J$ so that 
\re{nus} holds only for $J\gg \xi^2$. For $J=0$ the  first relation in \re{nus} reads
\begin{align}\label{nu0}
\nu_2(0) = -\xi ^2+\frac{\xi ^6}{2}+O\left(\xi ^{10}\right)\,.
\end{align}
In distinction to \re{nus}, the weak-coupling expansion of $\nu_2(0)$ involves only odd powers of $\xi^2$. In the next subsection, we exploit this
property to determine the $\xi^2-$odd part of the amplitude \re{A-plus}.
As was already mentioned, for
the integral in \re{A4-int} to be well-defined, $\xi^2$ should have a nonzero imaginary part. For $\Im \xi^2>0$ the pole \re{nu0} is located in
the lower half-plane.

Deforming the integration contour in  \re{A4-int} to the lower half-plane, we pick up the residue at the poles $\nu_2(J)$ and $\nu_4(J)$ to obtain 
\begin{align}\label{A4-sum} 
A(z,\xi^2)  {}&=- 16i \pi^2 \sum_{J \ge 0}\, \sum_{\nu=\nu_2(J),\nu_4(J)} \frac{2^{1-J} (J+1)\left((J+1)^2+4 \nu ^2\right)\nu }{(J+1)^2+4 \nu ^2+1} \, \Omega_{\nu,J} (z) [1+(-1)^J]\,.
\end{align}
Here, in a close analogy to \re{G-cont}, the sum runs over the states with an arbitrary (even) Lorentz spin $J$ and the scaling dimension $\Delta_2=2+2i\nu_2$ and
$\Delta_4=2+2i\nu_4$. 

There is however an important difference between \re{A4-sum} and \re{G-cont}. Arriving at \re{A4-sum} we have interchanged the sum over
spins with the integration over $\nu$ and, then, neglected the contribution from large $\nu$. In the case of the correlation function \re{G-cont}, this is justified by
the fact that the conformal block $g_{\Delta,J}(u,v)$ suppresses the contribution from large $\Delta=2+2i\nu$. For the scattering amplitude \re{A4-sum},
the situation is different. The function $\Omega_{\nu,J} (z)$, being an analog of the conformal block for the scattering amplitude, grows exponentially
fast at large $\nu$ for any given spin $J$. As a consequence, the sum over spins on the right-hand side of \re{A4-sum} is not expected to be convergent for general $z$. The problem can be avoided by using the Watson-Sommerfeld representation \re{WS} instead. 

\subsection{Constant part of the amplitude}

The relation \re{A4-sum} can be used to determine the constant, $\xi^2-$odd part of the amplitude \re{A-plus}
\begin{align}\label{B-A}
A_-(\xi^2) = \frac12\left[A(z,\xi^2)- A(z,-\xi^2)\right].
\end{align}
 As was shown in Sect.~\ref{sect:weak}, $A_-(\xi^2)$ only receives contributions from factorizable diagrams.
The same diagrams (but in off-shell kinematics) also contribute to the four-point correlation function \re{G-cont}. 
They consist of a few irreducible subgraphs connected together through the double-trace vertices  (see e.g. \re{4loop}). 
These vertices are generated
by the double-trace interaction term $\tr(X^2)\tr(\bar X^2)$ which is given by the product of two conformal operators, $\tr(X^2)$ and $\tr(\bar X^2)$. As a consequence,
the contribution of such diagrams to the correlation function factorized in the planar limit into the product of three-point correlation functions $\vev{\dots \tr(X^2)}\vev{\tr(\bar X^2) \dots}$. In the OPE expansion \re{G-cont}, it contributes to the
partial wave with $J=0$. Going through the LSZ procedure, we expect that the constant part $A_-(\xi^2)$ should only arise from the $J=0$ term in
\re{A4-sum}.

The contribution of the states with $J=0$ to the scattering amplitude \re{A4-sum} is
\begin{align} \label{terms}
{}& A(z,\xi^2)\Big|_{J=0}  =  - 32i  \sum_{\nu=\nu_2(0),\,\nu_4(0)}   \frac{\left(4 \nu ^2+1\right) \sinh ^2(\pi  \nu )}{ \nu  \left(2 \nu ^2+1\right)}\,,
\end{align}
where we replaced $\Omega_{\nu,0} (z)$ by \re{Om0}. Here
 $\nu_2(0)$ and $\nu_4(0)$ are solutions to \re{h-eq} for $J=0$ satisfying $\Im \nu<0$, 
 \begin{align}\notag
{}& \nu_2(0)= -\frac{\sqrt{\sqrt{4 \xi ^4+1}-1}}{\sqrt{2}} = -  \xi ^2+\frac{\xi ^6}{2}-\frac{7 \xi ^{10}}{8}+O\left(\xi ^{14}\right) \,,
\\
{}& \nu_4(0)=- i\frac{\sqrt{\sqrt{4 \xi ^4+1}+1}}{\sqrt{2}}=-i-\frac{i \xi ^4}{2}+\frac{5 i \xi ^8}{8}+O\left(\xi ^{12}\right)\,,
\end{align}
where $\Im \xi^2>0$. Comparing the two expressions we observe that $\nu_2$ and $\nu_4$ are,  respectively, odd and even functions of $\xi^2$ at weak coupling.
As a result, between the two terms in \re{terms} only the one  with $\nu=\nu_2(0)$ contributes to \re{B-A}
 \begin{align} \label{B-res}
A_-(\xi^2)={}& i \frac{32\sqrt{2} \left(1-2 \sqrt{4 \xi ^4+1}\right) }{  \sqrt{4 \xi ^4+1} \sqrt{\sqrt{4 \xi ^4+1}-1}} \sinh ^2\left(\frac{\pi  \sqrt{\sqrt{4 \xi
   ^4+1}-1}}{\sqrt{2}}\right).
\end{align}
The second term in  \re{terms} with $\nu=\nu_4(0)$ contributes to the $\xi^2-$even part of the amplitude  \re{A-plus}. In a similar manner, we can verify using \re{P-grow} that
the terms on the right-hand side of \re{A4-sum} with $J>0$ only contribute to $A_+(z,\xi^2)$.

The relation \re{B-res} gives the exact expression for the $\xi^2-$odd part of the amplitude.
At weak coupling, it looks as
\begin{align}\notag
A_-= 32 i \pi^2 \bigg[{}& \xi ^2+\left(\frac{3}{2}+\frac{\pi ^2}{3}\right) \xi ^6+\left(-\frac{49}{8}+\frac{\pi
   ^2}{6}+\frac{2 \pi ^4}{45}\right) \xi ^{10}
\\
   {}&\  +\left(\frac{363}{16}-\frac{15 \pi
   ^2}{8}-\frac{\pi ^4}{45}+\frac{\pi ^6}{315}\right) \xi
   ^{14}+O(\xi^{18})\bigg]\,.
\end{align}
We verify that the first three terms inside the brackets are in  perfect agreement with the result of the explicit five-loop calculation
\re{B-loops}.  At strong coupling, for $\xi\gg1$, we find from \re{B-res} that the amplitude grows exponentially $A_-\sim -64\,i\e^{\pi\xi} / \xi$.

\section{High-energy limit at arbitrary coupling}

In this section, we extend the analysis of Sect.~\ref{sect:high} and determine the asymptotic behaviour of the
scattering amplitude \re{WS} in the high-energy limit $s_{13}\gg s_{12}$, or equivalently $z\to\infty$, for arbitrary coupling $\xi^2$. 
In this limit, we can replace $\Omega_{\nu,J}(z) $ in \re{WS} with its leading asymptotic behaviour \re{Om1}
\begin{align}\label{A-J}
A {}& =  8\pi  \int_C {dJ\over 2\pi i} {\pi \, z^J \over\sin(\pi J)} \int_{-\infty}^\infty d\nu   \frac{2^{1-J} (J+1) \left((J+1)^2+4 \nu ^2\right)\nu ^2\,Q_{\nu,J}  }{
   \left( \nu ^2+ {J^2}/{4}\right) \left(\nu ^2+ (J+2)^2/4\right)-\xi
   ^4 }   +(z\to -z)\,,
\end{align}
where $Q_{\nu,J}$ is given by \re{Q-coef} and the integration contour $C$ encircles the nonnegative integer $J$ in anti-clockwise direction.
Opening up the integration contour $C$ and deforming it to the left-half plane, we find that the singularity of the integrand at $J=J_0$ 
produces a contribution of the form $A\sim z^{J_0}$. The leading large$-z$ asymptotics of \re{A-J} comes from the right-most singularity with the maximal $\Re J_0$.

\subsection{Exact Regge trajectories}\label{sect:ex}

The integrand in \re{A-J} has two sets of Regge poles in the complex $J-$plane. The four poles come from the denominator in \re{A-J}. They satisfy \re{h-eq} and are located at
\begin{align}\notag\label{soln}
{}& J_2^\pm=-1 +\sqrt{1-4 \nu ^2\pm 4 \sqrt{\xi ^4-\nu ^2}} \,,
\\
{}& J_4^\pm=-1 -\sqrt{1-4 \nu ^2\pm 4 \sqrt{\xi ^4-\nu ^2}} \,,
\end{align}
so that $J_2^\pm + J_4^\pm=-2$.
In addition, there are poles at $J=-2-n\pm 2i\nu$ (with $n=0,1,\dots$) coming from the function $Q_{\nu,J}$ defined in  \re{Q-coef}. 
They are located to the left of the poles \re{soln} and produce a subleading contribution to the amplitude.

Viewed as functions of the scaling dimension $\Delta=2+2i\nu$ of the exchanged states,  $J_2^\pm$ and $J_4^\pm$ define 
four Regge trajectories in the complex $(\Delta,J)-$plane. They can be interpreted as different branches of the complex curve \re{h-eq}. 
An unusual property of the functions \re{soln}, reflecting the lack of unitarity in the fishnet theory, is that the Regge trajectories \re{soln}
collide in a pair-wise manner at $\nu^2=\xi^4$ and $\nu^2=-1/4\pm \xi^2$.
As we show below, the Regge trajectories \re{soln} describe both the high-energy asymptotic
behaviour of the scattering amplitudes and the scaling dimensions of the `physical' operators that enter the OPE expansion of the correlation functions \re{G-cont}.

As was already mentioned, the dominant contribution to \re{A-J} in the high-energy limit comes from the right-most Regge singularity. It is easy to see from  \re{h-eq} that
the maximal  $\Re(J)$ is achieved at $\nu=0$, or equivalently $\Delta=2$,
\begin{align}\label{J-R}
J_{R} =  \sqrt{1+4\xi^2}-1\,.
\end{align}
It belongs to the trajectory $J(\nu)\equiv J_2^+(\nu)$  
\begin{align}\label{lead}
J(\nu) = -1 +\sqrt{1-4 \nu ^2+4 \sqrt{\xi ^4-\nu ^2}}\,,
\end{align}
which is, therefore, the leading Regge trajectory.  The first subleading trajectory $J_2^-(\nu)$ has the intercept $J_2^-(0)= \sqrt{1-4\xi^2}-1$, it takes negative values for
$\xi^2>0$ and satisfies $J_2^-(0)< J_R$.
Evaluating the residue of \re{A-J} at $J=J(\nu)$ and replacing $Q_{\nu,J}$ with \re{Q-coef} we find
\begin{align}\label{A-mas}
A(z,\xi^2) = - 32\pi \int_{-\infty}^\infty  \frac{d\nu \, \nu  \sinh (2 \pi  \nu ) \Gamma (J-2 i \nu +2) \Gamma (J+2 i \nu
   +2)\, (z/2)^{J(\nu)}}{\sin(\pi J)\left(J (J+2)+4 \nu ^2\right) [\Gamma \left(\frac{J}{2}-i \nu +1\right) \Gamma
   \left(\frac{J}{2}+i \nu +1\right)]^2}    + (z\to -z)\,,
\end{align}
where $J=J(\nu)$ is given by \re{lead}. This relation describes the high-energy asymptotics of the amplitude for arbitrary coupling. 
We show in Sect.~\ref{sect:beyond} that, at weak coupling, the relation \re{A-mas} is in agreement with the five-loop calculation presented in Sect.~\ref{sect:high}.

The dependence on $\xi^2$ enters into \re{A-mas}
through the Regge trajectory \re{lead}.
At small  $\nu$, it scales as
\begin{align}\label{disp}
J(\nu) = J_R - \nu ^2 \frac{  \left(2 \xi ^2+1\right)}{\xi ^2 \sqrt{4 \xi ^2+1}}+ O\left(\nu ^4\right)\,.
\end{align}
As a consequence, the leading contribution to \re{A-mas} comes from the integration in the vicinity of $\nu=0$  
\begin{align}\label{A-R}
A(z,\xi^2) \sim \int d\nu\, \nu^2  z^{J(\nu)}\sim {z^{J_R}\over (\ln z)^{3/2}} \,,
\end{align}
where $J_R$ is given by \re{J-R}. Since $J_R>0$ for $\xi^2>0$, the amplitude \re{A-R} grows as $z\to\infty$.

The relation \re{A-R} holds in the high-energy limit $z\to\infty$ for arbitrary coupling $\xi^2$. At weak coupling, $J_R= \sqrt{1+4\xi^2}-1=2\xi^2 + O(\xi^4)$ and
the relation \re{A-R} agrees with the high-energy limit of the amplitude in the leading-logarithmic approximation  \re{A-regge}. At strong coupling, we find from
\re{J-R} that $J_R= 2\xi -1 + O(1/\xi)$, leading to $A\sim z^{2\xi}/(\ln z)^{3/2}$.

\subsection{Scaling dimensions}

The same equation $h(\nu,J)=\xi^4$ defines the scaling dimensions of the local operators \re{Delta} and the position of the Regge poles \re{soln}.
An important difference, however, is that the Lorentz spin $J$ is integer and $\nu$ is complex for the former, whereas $J$ is complex and $\nu$ is real for   the latter. 
This suggests that the scaling dimensions can be found by analytically continuing the Regge trajectories \re{soln} in the complex $\nu-$plane.

At weak coupling we get from \re{soln}  
\begin{align}\notag\label{tr-dim}
J_2^+=\Delta-2+{2\xi^4\over (\Delta-1)(\Delta-2)} + O(\xi^8)\,,
\\
J_4^- = \Delta-4-{2\xi^4\over (\Delta-2)(\Delta-3)} + O(\xi^8)\,,
\end{align} 
where  $\Delta=2+2i\nu$. Inverting these relations we find the corresponding scaling dimensions \cite{Gromov:2018hut}
\begin{align}\notag
{}& \Delta_2=2+J-{2\xi^4\over J(J+1)} + O(\xi^8)\,,
\\
{}& \Delta_4=4+J+{2\xi^4\over (J+2)(J+3)} + O(\xi^8)\,.
\end{align}
They describe the operators of twist two and four, respectively. 
The remaining two trajectories, $J_4^+$ and $J_2^-$, can be obtained from \re{tr-dim} by replacing $\Delta\to 2-\Delta$, as they correspond to shadow operators.  

\begin{figure}[h!]
 \centering
 \includegraphics[width = 125mm]{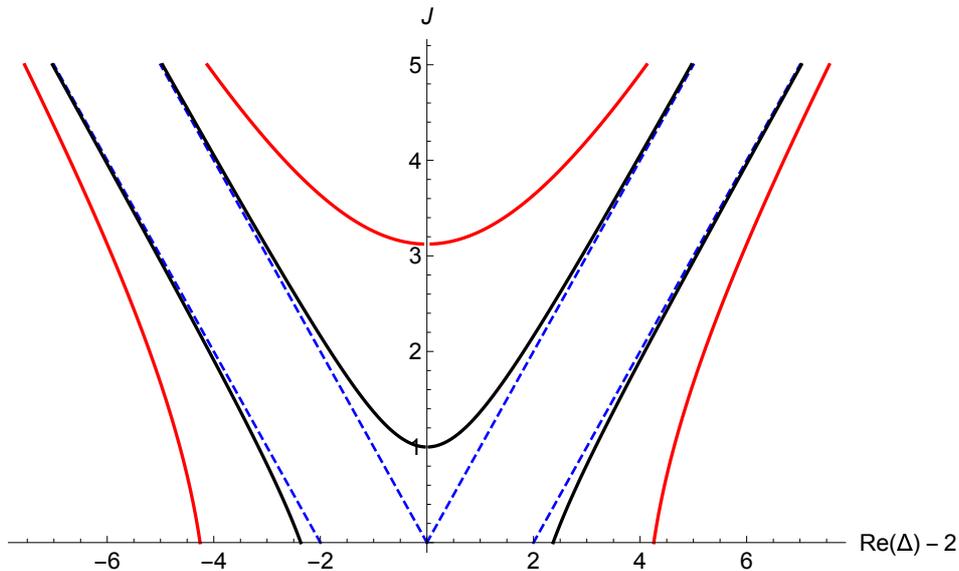}
\caption{The Regge trajectories for different values of the coupling: $\xi^2=0$ (dashed lines),   $\xi^2=0.75$ (black lines) and $\xi^2=4$ (red lines). The scaling dimensions
of local operators correspond to nonnegative even $J$ and satisfy $\Re \Delta\ge 2$.} 
\label{fig:traj}
\end{figure} 

The Regge trajectories \re{soln} are shown in Fig.~\ref{fig:traj}. 
The local conformal operators carry nonnegative even spin $J$ and their scaling dimensions satisfy the condition $\Re \Delta\ge 2$.~\footnote{Because the theory is non unitary, scaling dimensions can take complex values~\cite{Korchemsky:2015cyx}.}  For given coupling $\xi^2$, the upper and lower trajectories of the same color in Fig.~\ref{fig:traj} describe operators with twist two and twist four, respectively. The former trajectory crosses the line $\Delta=2$ at 
$J=J_R$  with $J_R$ being the leading Regge singularity \re{J-R}.  For $J<J_R$  the scaling dimension of twist-two operators develops a
square-root branch cut $\Delta_2-2\sim \sqrt{J-J_R}$. 

\subsection{Leading logarithmic approximation and beyond}\label{sect:beyond}

We have shown in Sect.~\ref{sect:high} that the perturbative corrections to the scattering amplitude at weak coupling are enhanced by powers of $\ln z$. Such corrections can be organized by 
considering the limit $L=\xi^2 \ln (z/2)=\text{fixed}$ as $\xi^2\to 0$. In this limit, the amplitude is given by a series in $\xi^2$ with the coefficients depending on $L$.
The first term of the expansion corresponds to the leading logarithmic approximation \re{A-lla}.
In this subsection, we apply \re{A-J} to reproduce \re{A-lla} and to systematically derive the subleading logarithmically enhanced terms.

In the high-energy limit $z\to\infty$, we shift the integration contour over $J$ in \re{A-J} to the left and pick up the residues at the Regge poles \re{soln}. Then, the logarithmically enhanced terms arise from the expansion of
$z^{J_t^\pm}$ (with $t=2,4$) at weak coupling. We verify that for $\nu=O(\xi^2)$, the Regge trajectories \re{soln} scale as $J_2^\pm=O(\xi^2)$ and $J_4^\pm=-2+O(\xi^2)$. 
Therefore, the leading contribution to the amplitude comes only from the twist-two trajectories $J_2^\pm$. Notice that the functions $J_2^+$ and $J_2^-$ coincide
at $\nu^2=\xi^4$ and develop a square-root branch cut for $\nu^2>\xi^4$. The cut disappears in the sum of the contributions of the two trajectories, so that the amplitude 
is analytic in $\nu$. Deforming the integration contour over $\nu$ in \re{A-J}, we can obtain after some algebra 
\begin{align}\label{A-fin}
{}& A(z,\xi^2)=  \int_{-\xi^2}^{\xi^2} d\nu \, \left[ F(\nu,J_+) (z/2)^{J_+}-F(\nu,J_-)  (z/2)^{J_-}  \right] + (z\to -z)\,,
\end{align}
where $J_\pm\equiv J_2^\pm(\nu)$ is the Regge trajectory \re{soln} and we introduced
\begin{align}
F(\nu,J) =- \frac{32\pi   \nu  \sinh (2 \pi  \nu ) \Gamma (J-2 i \nu +2) \Gamma (J+2 i \nu
   +2)}{\sin(\pi J)\left(J (J+2)+4 \nu ^2\right) [\Gamma \left( {J}/{2}-i \nu +1\right) \Gamma
   \left( {J}/{2}+i \nu +1\right)]^2}  \,.
\end{align}
We would like to emphasize that the relation \re{A-fin} describes all the logarithmically enhanced contributions to the amplitude of the form $(\xi^{2})^{k+1} (\ln z)^n$ (with $n\le k$)
and it holds up to corrections suppressed by powers of $1/z$.   In comparison with \re{A-mas},  the integration 
in \re{A-fin} goes over a finite interval $\nu^2 <\xi^4$.
Notice that, in distinction to \re{A-R}, the relation \re{A-fin} receives contributions from both twist-two trajectories. Indeed, 
for $L=\xi^2 \ln (z/2)=\text{fixed}$ and $\xi^2\to 0$
both terms $z^{j_+}$ and $z^{j_-}$ generate powers of $L$, whereas in the Regge limit, for $z\to\infty$ and $\xi^2=\text{fixed}$, the latter is suppressed.

It is convenient to change the integration variable in \re{A-fin} to $\nu=\xi^2\sqrt{1-x^2}$. Then, introducing the notation for
\begin{align}\notag
j(x) {}& =\lr{\sqrt{1+4 \xi ^2 x+4 \xi ^4 \left(x^2-1\right)}-1 }/\xi^2
\,,
\\[2mm]
f(x) {}& = F(\xi^2\sqrt{1-x^2},\xi^2 j(x)) \,,
\end{align}
we find from  \re{A-fin} 
\begin{align} \notag \label{A-integ}
A(z,\xi^2) {}&= 2\xi^2 \int_{0}^1 {dx \, x\over \sqrt{1-x^2}} [f(x) \e^{L j(x)} - f(-x) \e^{L j(-x)}]+ (z\to -z)
\\ \notag
{}& = 2\xi^2 \int_{-1}^1 {dx \, x\over \sqrt{1-x^2}} f(x) \e^{L j(x)} + (z\to -z)
\\
{}& =  i\xi^2 \oint_{[-1,1]} {dx \, x\over \sqrt{x^2-1}} f(x) \e^{L j(x)} + (z\to -z)
\,,
\end{align}
where $L=\xi^2 \ln (z/2)$ and the integration contour in the last relation encircles the interval $[-1,1]$ in an anticlockwise direction. 

We can apply \re{A-integ} to determine the logarithmically enhanced  corrections to the scattering amplitude at weak coupling to any order in  $\xi^2$. The integral in \re{A-integ} can be easily evaluated by residues.
A close examination shows that at small $\xi^2$ the integrand in \re{A-integ} has poles at 
$x=0$ and $x\to\infty$. The former pole arises due to $xf(x)\sim 1/(x+\xi^2)$ as $\xi^2\to0$. Because $\xi^2$ has a nonzero imaginary part, it
is located  outside the integration contour. Blowing up the integration contour in \re{A-integ}, we find 
\begin{align}
A(z,\xi^2) = A_+ + A_- + (z\to -z)\,,
\end{align}
where $A_+$ and $A_-$ are given by the residues at $x=0$ and $x\to\infty$, respectively,
\begin{align}\notag
{}& A_+ =   
16i\pi^2 \left[ \xi^2 + \xi^6 \left(\frac32+ \frac{\pi
   ^2}3\right)  + \xi^{10}\left(-\frac{49}{8}+\frac{\pi
   ^2}{6}+\frac{2 \pi ^4}{45}\right) +O\left(\xi^{14}\right)\right],
\\  
{}& A_- = 16\pi^2\bigg[  -(\ell+1) \xi ^4-  \left(\frac{\ell^3}{6}+\frac{\pi ^2}{2} \ell +  4 \zeta_3-3 \right) \xi ^8  -\bigg(\frac{\ell^5}{60}-\frac{\ell^4}{12}+\frac{\pi ^2 }{9}\ell^3 
   \\\notag
{}&  
  + \Big(\zeta_3-\frac{\pi
   ^2}{3}\Big)\ell^2 + \Big(\frac{17 \pi ^4}{180}-2 \zeta_3\Big)\ell -12 \zeta_5+\frac{5
   \pi ^2 \zeta_3}{3}-\frac{17 \pi ^4}{180}-\frac{2 \pi ^2}{3}+12\bigg) \xi ^{12}+O\left(\xi^{16}\right)\bigg],
\end{align}
with $\ell=\ln (z/2)$. We verify that these relations are in  perfect agreement with the result of the five-loop calculation \re{B-loops} and \re{A-regge2}.

As the  next step, we expand the integrand of \re{A-integ} in powers of $\xi^2$  with $L=\xi^2 \ln (z/2)=\text{fixed}$ to obtain
\begin{align}
A_-(z,\xi^2)= - 16\pi^2 \left[ \xi^2 A_{\rm LLA} (L) + \xi^4 A_{\rm NLA} (L)+ \xi^6 A_{\rm N^2LA} (L)+ O(\xi^8)\right]\,,
\end{align}
where the first term on the right-hand side describes the leading logarithmic approximation, the second one is the next-to-leading approximation etc. The functions
$A_{\rm N^kLA}(L)$ take the following form
\begin{align}\label{Ak}
A_{\rm N^kLA} = {1\over\pi} \dashint_{-1}^1 {dx\over x^{k+1}} \sqrt{1-x^2} \e^{2 L x} a_{2k}(x,L)+ (z\to -z)\,,
\end{align}
where the integral is defined using the principal value prescription and $a_{2k}(x,L)$ are polynomials in $x$ of degree $2k$ with coefficients depending on $L$ 
\begin{align}\notag\label{ak}
{}& a_0 = 1\,,\qqqquad 
\\[2mm] \notag
{}& a_2= 4x^2-2Lx +1\,, 
\\
{}& a_4=2 L^2 x^2-2 L \left(2 x^3+x\right)+\frac{1}{3} \left(2 x^2+1\right) \left(\pi ^2 x^2+3\right)\,, \quad \dots
\end{align}
Substituting the first relation into \re{Ak} we arrive at \re{A-lla}. As before, the integral in \re{Ak} can be evaluated by converting it into a contour integral encircling the interval $[-1,1]$ and taking the residue at infinity. In this way, we apply \re{Ak} and \re{ak} to derive higher order corrections to the scattering 
amplitude, e.g. (see footnote~\ref{foot})
\begin{align}\notag\label{a,b-sub}
{}& A_{\rm NLA}= \sum_{n\ge 0}L^{2n} {  (n-1)\over n! (n+1)!}\,,
\\
{}& A_{\rm N^2LA}= - \sum_{n\ge 0}L^{2 n+1} \frac{\left(2  n(n-1)(n+2)+\pi ^2 (n+1)\right) }{(2 n+1)  n! (n+2)!}\,,\quad \dots
\end{align}
It would be challenging to reproduce these relations by a direct calculation of the Feynman diagrams shown in Fig.~\ref{fig:xx}.

\section{Conclusions}

In this paper, we have computed four-particle scattering amplitudes in the conformal fishnet theory. This theory arises as a special limit
of the $\gamma-$deformed $\mathcal N=4$ SYM and it inherits the remarkable integrability properties of  the latter theory. In distinction to 
 $\mathcal N=4$ SYM, the four-particle amplitudes in the fishnet theory are free from infrared divergences in the leading large-$N$ limit and enjoy (unbroken) conformal symmetry.  
The single-trace amplitude is protected from quantum corrections in the planar limit whereas the double-trace amplitude is a nontrivial function of a single variable given by the ratio of the independent Mandelstam invariants. At weak coupling,
we computed this function at five loops by applying the  conventional Feynman diagram technique. We demonstrated that, in the high-energy limit, the double-trace amplitude has a Regge like asymptotic bevaviour and computed the corresponding leading Regge trajectory.

The main advantage of the fishnet theory as compared with $\mathcal N=4$ SYM is that, due to the  particular  (chiral) form of the quartic scalar interaction, it allows for finding the exact expression for the four-point correlation
function of the scalar fields in the leading large-$N$ limit. Applying the LSZ reduction formula to this correlation function, we derived 
a new representation for the double-trace amplitude \re{amp-0} as a sum over conformal partial waves. It follows from the  analogous expansion of the correlation function over the conformal blocks and involves a new ingredient - the conformal polynomial \re{P-grow}.
We applied this representation to find the exact expression for the $\xi^2-$odd part of the double-trace amplitude \re{B-res}. 
For the $\xi^2-$even part of the amplitude, we examined its asymptotic behavior in the high-energy limit and found the exact expressions for the corresponding Regge trajectories. At weak coupling, the expressions obtained are in  perfect agreement with the result of the five-loop calculation. At strong coupling, the leading Regge singularity scales as $O(\sqrt{\xi^2})$. It would be interesting to reproduce the same
behaviour using the  dual description of the conformal fishnet theory \cite{Basso:2018agi}.

The representation \re{amp-0} relies on  conformal symmetry, and should be applicable to the four-particle amplitudes in $\mathcal N=4$ SYM beyond the planar limit. 
More precisely, the latter amplitudes suffer from IR divergences and satisfy anomalous conformal Ward identities.  The  homogenous solution to these identities should admit the representation similar to \re{amp-0}. It may also shed light on the 
properties of nonplanar amplitudes in $\mathcal N=4$ SYM. It would be interesting to apply \re{amp-0} to the three-loop result for 
the four-gluon scattering amplitude in $\mathcal N=4$ SYM derived in Ref.~\cite{Henn:2016jdu}.

It would also be interesting to extend the above consideration to higher-point amplitudes in the fishnet theory. Due to the  nonzero total $U(1)\times U(1)$ charge, the amplitudes with an odd number of scalars vanish. The simplest six-point amplitude is of special interest 
- the analogous amplitude in planar $\mathcal N=4$ SYM is dual to a hexagon light-like (super) Wilson loop and has a number of remarkable properties. In the fishnet theory, the planar six-particle amplitude is given by a single tree-level diagram \cite{Chicherin:2017cns,Chicherin:2017frs}. As a consequence, similar to the four-particle case, the single-trace contribution to the six-particle amplitude is protected from quantum corrections in the planar limit. 
The leading-color contribution to the double- and triple-trace partial amplitudes can be obtained by applying the LSZ reduction
procedure to a six-point correlation function of scalar fields. In the fishnet theory, this correlation function can be expanded over the conformal
partial waves in different OPE channels.

\section*{Acknowledgements} 

I would like to thank Volodya Kazakov for collaboration at the early stage of this project.
I am grateful to Volodya Kazakov, David Kosower and Emeri Sokatchev for useful discussions and helpful comments.. This work was supported by the French National Agency for Research grant
ANR-17-CE31-0001-01. I would like to thank the Galileo Galilei Institute for Theoretical Physics
for its hospitality, and INFN and Simons Foundation for partial support during the completion of this
work.

\appendix

\section{Conformal basis in momentum space}\label{app:A}

In this appendix we apply the LSZ reduction formula \re{Phi-F1} and derive \re{Phi-F}. 
The relation \re{Phi-F1} involves the function 
$\Phi_{\nu,J} (x_{10},x_{20})$ introduced in \re{Phi}. It can be identified as the three-point correlation function 
\begin{align}
\Phi_{\nu,J}(x_{10},x_{20}) =\vev{X(x_1) X(x_2) O_{\Delta,J}(x_0)}\,,
\end{align}
where $O_{\Delta,J}(0)$ is the primary operator with the scaling dimension $\Delta=2+2i\nu$ and Lorentz spin $J$. Its Fourier transform
defines the off-shell form factor
\begin{align} \label{ff}
F_{\nu,J}(p_1,p_2) {}&= \int d^4 x_1 d^4 x_2 \e^{ip_1 x_1+ip_2 x_2} \Phi_{\nu,J}(x_1,x_2) =\sum_{k=0}^J (-1)^{J-k} \lr{J\atop k} I_{k,J-k} \,,
\end{align}
where $p_i^2\neq 0$ and the notation was introduced for 
\begin{align} 
I_{k_1,k_2}
{}&=\int {d^4 x_1 d^4 x_2 \e^{ip_1 x_1+ip_2 x_2}  \over (x_{12}^2)^{1-t/2}(x_1^2x_2^2)^{t/2}} \left[{{2(nx_1)\over x_1^2}}\right]^{k_1}\left[{{2(nx_2)\over x_2^2}}\right]^{k_2},
\end{align}
with $t=\Delta-J$ and $J=k_1+k_2$. 

We expect that for $p_1^2,p_2^2\to 0$  the integral develops a double pole $ I_{k_1,k_2}\sim 1/(p_1^2 p_2^2)$.
Indeed,  $ I_{k_1,k_2}$ admits the Mellin-Barnes representation 
\begin{align} \notag
I_{k_1,k_2} {}&= (in{p_1})^{k_1}(in{p_2})^{k_2} c_{k_1 k_2}
 \int {dj_1 dj_2\over (2\pi i)^2}  \lr{p_1^2\over 4}^{j_1}\lr{p_2^2\over 4}^{j_2}
{(-s_{12}/4)^{-j_1-j_2+ {t}/{2}-3}\over \Gamma \left(J+{t}/{2}-1\right)}\Gamma(-j_1)\Gamma(-j_2)
\\  
{}& \times  
{\Gamma \left(-j_1-1\right) \Gamma \left(-j_2-1\right) \Gamma
   \left(-{t}/{2}+j_1+j_2+3\right)   \Gamma
   \left(J+{t}/{2}+j_1+j_2+1\right)},
\end{align}
where $c_{k_1 k_2}={\pi^4/(\Gamma(1-t/2)\Gamma(t/2+k_1)\Gamma(t/2+k_2))}$ and $s_{12}= (p_1+p_2)^2$.
The double pole $1/(p_1^2p_2^2)$ arises as the contribution of the poles at $j_1=j_2=-1$
\begin{align}
I_{k_1,k_2} {}&={(2\pi)^4\over p_1^2p_2^2} (-s_{12}/4)^{\frac{t}{2}-1}  { (in{p_1})^{k_1}(in{p_2})^{k_2}\over \Gamma(t/2+k_1)\Gamma(t/2+k_2)}
+\dots
\end{align}
where the dots denote subleading terms. Together with \re{ff} this leads to
\begin{align}
F_{\nu,J}(p_1,p_2) {}&={(2\pi)^4\over p_1^2p_2^2} \widetilde \Phi_{\nu,J}(p_1,p_2) +\dots
\end{align}
where $\widetilde \Phi_{\nu,J}(p_1,p_2)$ is given by  
\begin{align}\label{on-shell}
 \widetilde \Phi_{\nu,J} (p_1,p_2) {}&=(-s_{12}/4)^{\frac{t}{2}-1}  \sum_{k=0}^J (-1)^{J-k}   \lr{J \atop k}  { (in{p_1})^{k}(in{p_2})^{J-k}\over \Gamma(t/2+k)\Gamma(t/2+J-k)} \,,
\end{align}
with $t=\Delta-J=2+2i\nu-J$. The sum on the right-hand side of \re{on-shell} can be expressed in terms of Gegenbauer polynomials
leading to \re{Phi-F}. 
The function \re{on-shell} has the meaning of an on-shell form factor, $ \widetilde \Phi_{\nu,J} (p_1,p_2)=\vev{p_1,p_2|O_{\Delta,J}(0) |0} $.

As a check, we examine \re{on-shell} for  $\nu=-iJ/2$, or equivalently for $\Delta=2+J$   
\begin{align}\label{check}
\widetilde \Phi_{\nu=-iJ/2,J}(p_1,p_2)
={ i^J\over J!}  (\xi_1+\xi_2)^J P_J\left(\xi_1-\xi_2 \over \xi_1+\xi_2\right)\,.
\end{align}
The corresponding conformal operator $O_{\Delta,J}$ with  scaling dimension $\Delta=2+J$ can be constructed in the  free theory from two scalar fields and $J$
light-cone derivatives. 
It takes the well-known form (see e.g. Ref.~\cite{Braun:2003rp})
\begin{align}
O_{\Delta=J+2,J}(0) = {1\over J!}  (\partial_1+\partial_2)^J P_J\left(\partial_2-\partial_1 \over \partial_2+\partial_1\right) \bar X(x_1) \bar X(x_2) \Big|_{x_1=x_2=0}\,,
\end{align}
where $\partial_i = (n\partial_{x_i})$.  It is easy to see that its on-shell matrix element is given by \re{check}.

\section{Conformal polynomial}\label{app:B}

To compute the function $\Omega_{\nu,J}(z)$ defined in \re{P-def}, we apply \re{D-op} and \re{on-shell} to construct two completely
symmetric traceless tensors, $\widetilde \Phi_{\nu}^{\mu_1\dots\mu_J}(p_1,p_2)$ and $\widetilde \Phi_{-\nu}^{\mu_1\dots\mu_J}(p_3,p_4)$, 
and, then, substitute them into \re{P-def}. 
For $J=0$ we have
\begin{align}\label{Om0}
\Omega_{\nu,0}(z) =   \widetilde \Phi_{\nu,0}(p_1,p_2)\widetilde \Phi_{-\nu,0}(p_3,p_4) = {\sinh^2(\pi\nu)\over (\pi\nu)^2}\,.
\end{align}
For $J\ge 1$, the function $\Omega_{\nu,J}(z)$ has the  general form \re{Om1}. 

We can find the leading term in \re{Om1} by considering \re{P-def} in the limit $p_1\to -p_2$ and $p_3\to -p_4$, or equivalently $s_{12}\to 0$ and $s_{13}=\rm{fixed}$.
In this limit $z\to\infty$ and $\Omega_{\nu,J}(z) \sim Q_{\nu,J}z^J$. 
For $p_2\to -p_1$ we can safely replace $(np_2)\to -(np_1)$ in \re{on-shell} to get
\begin{align} 
\widetilde \Phi_{\nu}^{\mu_1\dots\mu_J}(p_1,p_2){}&=  i^J (-s_{12}/4)^{i\nu-J/2}\frac{2^{J+2 i \nu } \Gamma \left(\frac{1}{2} (J+2 i \nu +1)\right)}{\sqrt{\pi } \Gamma
   (2 i \nu +1) \Gamma \left(\frac{J}{2}+i \nu +1\right)}p_1^{\mu_1}\dots p_1^{\mu_J}+\dots
\end{align}
and similar for  $\widetilde \Phi_{-\nu}^{\mu_1\dots\mu_J}(p_3,p_4)$. Then, we use this relation to find from \re{P-def}
\begin{align} 
\Omega_{\nu,J}(z) {}&= z^J\frac{\sinh (2 \pi  \nu ) \Gamma (J-2 i \nu +1) \Gamma (J+2 i \nu +1)}{2 \pi  \nu  [\Gamma
   \left(\frac{J}{2}-i \nu +1\right) \Gamma \left(\frac{J}{2}+i \nu +1\right)]^2} +\dots\,,
\end{align}
where $z\sim 2s_{13}/s_{12}$. The coefficient in front of $z^J$ can be identified as $Q_{\nu,J}$, see Eq.~\re{Q-coef}.

To find $\Omega_{\nu,J}(z)$ for arbitrary $z$, we introduce the  completely symmetric traceless tensor 
$T_k^{\mu_1\dots \mu_J} (p_1,p_2)$ satisfying the defining relation
\begin{align}\label{T-def}
T_k^{\mu_1\dots \mu_J} (p_1,p_2) n_{\mu_1}\dots n_{\mu_J} = (-1)^{J-k}   (np_1)^k (np_2)^{J-k} \lr{J\atop k}\,,
\end{align}
where $p_i^2=0$ and $n_\mu$ is an auxiliary light-like vector.
Then, we obtain from \re{on-shell} and \re{D-op}   
\begin{align}
\widetilde \Phi_{\nu}^{\mu_1\dots\mu_J}(p_1,p_2) = i^J (-s_{12}/4)^{\frac{t}{2}-1}  \sum_{k=0}^J  { T_k^{\mu_1\dots \mu_J} (p_1,p_2) \over \Gamma(t/2+k)\Gamma(t/2+J-k)} \,.
\end{align}
Substituting this relation into \re{P-def} we get
\begin{align}\label{TT}
\Omega_{\nu,J}(z) {}&=(s_{12}/4)^{-J}
\sum_{k,m =0}^J  { T_k^{\mu_1\dots \mu_J} (p_1,p_2) T_m^{\mu_1\dots \mu_J} (p_3,p_4) \over \Gamma(t/2+k)\Gamma(t/2+J-k)  \Gamma(\bar t/2+m)\Gamma(\bar t/2+J-m)} \,,
\end{align}
where $t=\Delta-J=2-J + 2i\nu$ and $\bar t =2-J - 2i\nu$. 

To evaluate the product of two tensors in the numerator of \re{TT} we apply the 
identity
\begin{align} 
f_J(k_1,k_2) 
 {}&= (k_1^{\mu_1} \dots k_1^{\mu_J} -\text{traces}) (k_2^{\mu_1} \dots k_2^{\mu_J} -\text{traces})= \lr{k_1^2 k_2^2\over 4}^{J/2} C_J^1\left(\cosh\theta  \right),
\end{align}
where $C_J^1$ is the Gegenbauer polynomial and $\cosh\theta = {(k_1 k_2)/ (k_1^2 k_2^2)^{1/2}}$.
Replacing $k_1=z_1 p_1 - p_2$ and $k_2=z_3p_3-p_4$ we can expand $f_J(k_1,k_2)$  in powers of $z_i$. As follows from \re{T-def}, the corresponding
expansion coefficients are given by the product of two tensors that appears in \re{TT}
\begin{align}
f_J(z_1 p_1 - p_2,z_3p_3- p_4)= \sum_{k,m =0}^J  z_1^k  z_3^m  \,
T_k^{\mu_1\dots \mu_J} (p_1,p_2) T_m^{\mu_1\dots \mu_J} (p_3,p_4) \,.
\end{align}
Matching the expressions on the right-hand side of the last two relations, we can find $T_k^{\mu_1\dots \mu_J} (p_1,p_2) T_m^{\mu_1\dots \mu_J} (p_3,p_4)$
and, then, evaluate \re{TT} for any given $J$. In this way we obtain 
\begin{align}\notag
\Omega_{\nu,1} {}&= -\frac{4 z \cosh ^2(\pi  \nu )}{\pi ^2 \left(\nu ^2+1/4\right)}\,,
\\[2mm]
\Omega_{\nu,2} {}&= \frac{4 \sinh ^2(\pi  \nu ) \left(4 \nu ^2 z^2+z^2-\nu ^2\right)}{\pi ^2 \nu ^2
   \left(\nu ^2+1\right)}\,.
\end{align}
For $J>2$ we found with some guesswork that the resulting expression for $\Omega_{\nu,J}(z)$ is given by \re{P-grow}.

Let us examine the properties of $\Omega_{\nu,J}(z)$  in the complex $\nu-$plane. 
Each term in the sum \re{P-grow} has a pole at $\nu=\pm i(J/2-k)$. It is compensated however by $\sinh^2(\pi \nu+i \pi J/2)$ so that 
 $\Omega_{\nu,J}(z)$ is an analytical function of $\nu$. 
At large $\nu$, we find from \re{P-grow} that $\Omega_{\nu,J}(z)$ grows exponentially fast
\begin{align}\label{Cheb}
 \Omega_{\nu,J}(z)/2^J \sim \frac{\sinh ^2 ( \pi  \nu)}{(\pi\nu)^2} \sum_{k=0}^J  P_k(z) P_{J-k}(z) 
= \frac{\sinh ^2 ( \pi  \nu)}{(\pi\nu)^2} U_J(z)\,,
\end{align}
where  $U_J(z)$ is  Chebyshev polynomial of the second kind.

\section{Leading logarithmic approximation}\label{app:C}

In the high-energy limit $z =1+2s_{13}/s_{12}\gg 1$, the scattering amplitude is enhanced by powers of logarithms (see Eqs.~\re{B-loops} and \re{A-regge2})
\begin{align}\label{a,b}
A= 16\pi^2 \sum_{n=0}^\infty \xi^{4n+4} \left[ a_{n} \ln ^{2n+1}(z/2) +b_{n} \ln^{2n} (z/2)  +  \dots \right].
\end{align}
In the leading logarithmic approximation, for $z\to\infty$ and  $L=\xi^2 \ln (z/2)$ fixed, we can retain terms proportional to
$a_{n}$.

To find the leading coefficients $a_{n}$, we examine the discontinuity of the amplitude with respect to $s_{13}$
\begin{align}\label{disc}
{\rm disc}_{s_{13}} A = 16\pi^2 \sum_{n=0}^\infty \xi^{4n+4} \left[ (2n+1)a_{n} \ln^{2n}(z/2)  +  \dots \right].
\end{align}
It is easy to see from \re{diags} that the discontinuity receives a contribution from ladder diagrams. The remaining factorizable diagrams 
involving double-trace vertices do not depend on $s_{13}$ and do not contribute to \re{disc}. In this way, we obtain 
\begin{align}\label{unit}
\psfrag{dots}[cc][cc]{$\mathbf{\vdots}$}\psfrag{s}[cc][cc]{$s$}
\psfrag{1}[cc][cc]{$1$}\psfrag{2}[cc][cc]{$2$}\psfrag{3}[cc][cc]{$3$}\psfrag{4}[cc][cc]{$4$}
\psfrag{l1}[cc][cc]{$\ell_1$}\psfrag{l2}[cc][cc]{$\ell_2$}\psfrag{l3}[cc][cc]{$\ell_n$}
\psfrag{m1}[lc][cc]{$q+\ell_1$}\psfrag{m2}[lc][cc]{$q+\ell_2$}\psfrag{m3}[lc][cc]{$q+\ell_n$}
 {\rm disc}_{s_{_{13}}} \, A \ =\ \sum_n \ \parbox[c]{55mm}{ \includegraphics[width = 55mm]{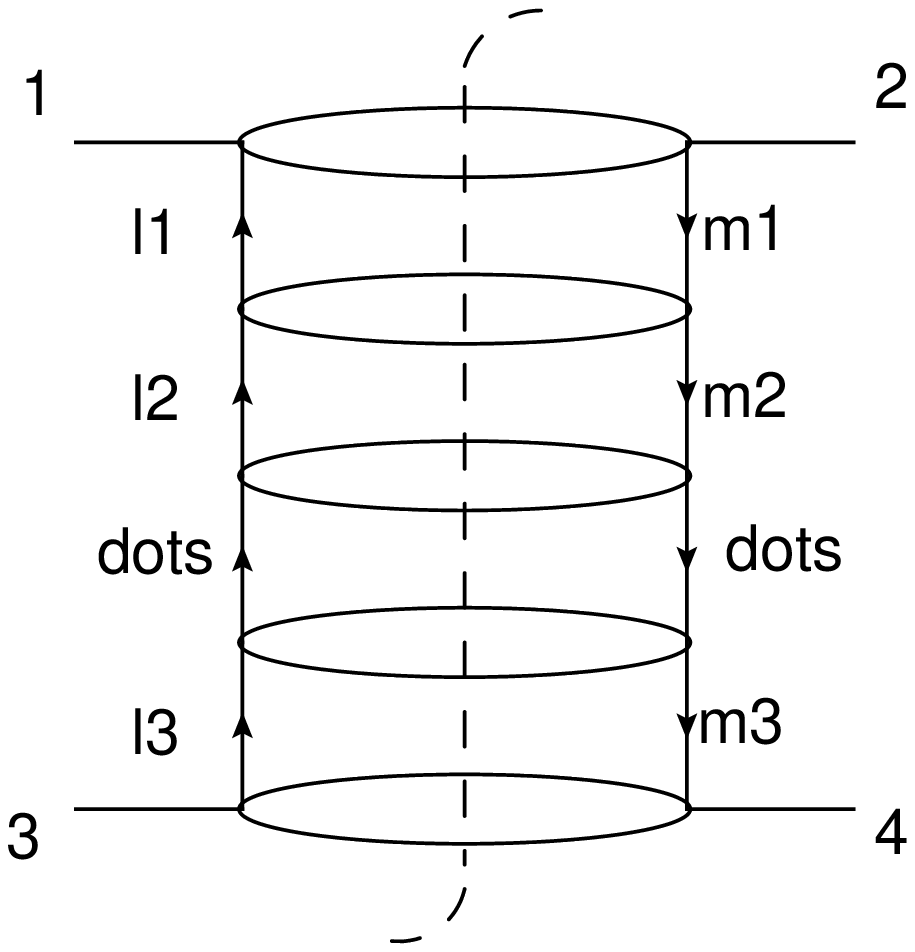}}
\end{align}
where the dashed line denotes the unitary cut and $q=p_1+p_2$ is the momentum transferred in the $t-$channel.

Computing the discontinuity we assume that $s_{13}>0$ and $s_{12}=q^2<0$ (we recall that all  the external on-shell momenta are incoming). The contribution of the diagram
in \re{unit} involves the Feynman integral
\begin{align}\label{int-cut}
\int \prod_{k=1}^n {d^4 \ell_k\over (2\pi)^4} {1\over \ell_k^2 (q+\ell_k)^2} \rho(\ell_1+p_1)\rho(\ell_2-\ell_1)\dots 
\rho(\ell_n-\ell_{n-1})\rho(p_3-\ell_n)\,,
\end{align}
which is finite in $D=4$ dimensions and does not require regularization.
Here $\rho(K)$ describes the cut scalar loop and $K$ is the total incoming momentum. It is given by the discontinuity of the function $\pi(s=K^2)$ defined in \re{pi} 
\begin{align}\label{pi}
\rho(K)={\rm disc}_{K^2}   \int {d^{4-2\epsilon} \ell \over i \, (2\pi)^{4-2\epsilon}} {1\over \ell^2 (K-\ell)^2} ={1\over 16\pi}\theta(K^2)\theta(K_0)\,.
\end{align}
The conditions $K^2>0$ and $K_0>0$ impose restrictions on the loop momenta in \re{int-cut}. 

To define  the integration region in \re{int-cut}
it is convenient to
introduce a Sudakov parameterization of the loop momenta
\begin{align}\label{Sudakov}
\ell_i = -\alpha_i p_1 + \beta_i p_3 + \ell_{i,\perp}\,,  \qqqquad \int d^4 \ell_i = s_{13} \int d\alpha \, d\beta \,  d^2 \ell_{i,\perp}\,,
\end{align}
where $\ell_{i,\perp}$ is a two-dimensional Euclidean vector orthogonal to $p_1$ and $p_3$.  The following relations hold
\begin{align}\notag
{}& \ell_i^2 = -s_{13} \alpha_i \beta_i -  {\vec \ell}_{i,\perp}^2\,, &&   (\ell_i-\ell_j)^2 =- s_{13}(\alpha_i-\alpha_j)(\beta_j-\beta_i) - {\vec \ell}_{ij,\perp}^2\,,
 \\[2mm]
{}&  (p_1+\ell_1)^2 = s_{13}(1-\alpha_1)\beta_1 - {\vec \ell}_{1,\perp}^2\,,
 && (p_3-\ell_n)^2 = s_{13}\alpha_n(1- \beta_n) - {\vec \ell}_{n,\perp}^2\,,
\end{align}
where ${\vec \ell}_{ij,\perp}^2 =({\vec \ell}_{i,\perp}-{\vec \ell}_{j,\perp})^2$.
The integration over the Sudakov variables in \re{Sudakov} is restricted to the region $1>\alpha_1>\dots > \alpha_n>0$ and $1>\beta_n>\dots > \beta_1>0$ subject to the conditions $(\ell_i-\ell_j)^2>0$, $(p_1+\ell_1)^2>0$ and $(p_3-\ell_n)^2>0$.

An additional simplification arises in the high-energy limit $s_{13}\gg s_{12}$. In this limit, in the leading logarithmic approximation, the dominant contribution to \re{int-cut} comes from the integration over the strongly ordered Sudakov
variables \cite{Gribov:2003nw}
\begin{align} \label{order}
{}& s_{12}/s_{13} \ll \alpha_n \ll  \dots \ll \alpha_1 \ll 1
\,,\qqqquad s_{12}/s_{13} \ll \beta_1 \ll   \dots  \ll \beta_n \ll 1\,,
\end{align}
and $\vec l_{i,\perp}^2=O(s_{12})$. Since $q^2=s_{12} \ll s_{13}$ in the high-energy limit, we can safely replace the scalar propagators $(q+\ell_k)^2$ in \re{int-cut} with $\ell_k^2$. Then, the integration over the transverse momenta $\ell_{i,\perp}$ in \re{int-cut} yields
\begin{align}
\int {d\alpha_1 d\beta_1\over \alpha_1 \beta_1 + m^2/s_{13}}\dots \int {d\alpha_n d\beta_n\over \alpha_n \beta_n + m^2/s_{13}}\,,
\end{align}
where $\alpha_i$ and $\beta_i$ satisfy \re{order} and  $m^2=O(s_{12})$ plays the role of an IR cut-off. The integral can be evaluated using
the Mellin-Barnes representation
\begin{align}\notag
{}&  \int_{-\delta-i\infty}^{-\delta+i\infty} {dj_1\over 2\pi i}{\pi\over \sin(\pi j_1)} 
\dots {dj_N\over 2\pi i} {\pi\over \sin(\pi j_n)}  (m^2/s_{13})^{j_1+\dots+j_n}
\\
{}&\times
\int_0^1 d\alpha_1 \alpha_1^{-1-j_1} \dots \int_0^{\alpha_{n-1}} d\alpha_n \alpha_n^{-1-j_n} 
\int_0^1 d\beta_n \beta_n^{-1-j_n} \dots \int_0^{\beta_{2}} d\beta_1 \beta_1^{-1-j_1} 
\end{align}
Closing the integration contour to the right-half plane and evaluating the residue at $j_i=0$ we find the leading asymtotic behavior for $m^2/s_{13}\to 0$ as $\ln^{2n} (s_{13}/m^2)/(n!(n+1)!)$. Comparing with the expression inside the brackets in \re{disc} we deduce that
\begin{align}
a_{n} = {1\over (2n+1)n! (n+1)!}  \,.
\end{align}
To find the subleading coefficients in \re{a,b}, one has to relax the condition of strong ordering of the Sudakov variables \re{order}. The
expressions for these coefficients can be read from \re{a,b-sub}.

\bibliographystyle{JHEP} 


\providecommand{\href}[2]{#2}\begingroup\raggedright\endgroup

\end{document}